\documentclass[11pt,letterpaper]{aastex63}

\usepackage{bm}
\usepackage{amssymb}
\usepackage{amsmath}
\newcommand{\doctype}{paper}
\newcommand{\equationname}{equation}
\newcommand{\eqnref}[1]{\mbox{\equationname~\ref{#1}}}
\newcommand{\appref}[1]{\mbox{Appendix~\ref{#1}}}

\newcommand{\niceurl}[1]{\mbox{\href{#1}{\textsl{#1}}}}
\newcommand{\latin}[1]{\emph{#1}}

\newcommand{\ie}{\latin{i.e.}}

\newcommand{\figref}[1]{\mbox{Figure~\ref{#1}}}
\newcommand{\detmap}{detection map}

\newcommand{\drawnfrom}{\sim}
\newcommand{\gaussianN}{\mathcal{N}}
\newcommand{\gaussian}[1]{\gaussianN\!\left(#1\right)}
\newcommand{\gaussx}[2]{\gaussianN\!\left(#1 \, , \, #2\right)}
\newcommand{\psf}{\psi}
\newcommand{\psfat}[1]{\psf_{#1}}
\newcommand{\psfnorm}{\norm{\bm{\psf}}}

\newcommand{\snr}[1]{\mathbb{SN}(#1)}

\newcommand{\norm}[1]{\left\lVert #1 \right\rVert}

\newcommand{\var}[1]{\mathrm{var}\left( #1 \right)}

\newcommand{\psfw}{w}
\newcommand{\dd}{\mathrm{d}}
\newcommand{\noise}{E}
\renewcommand{\vec}[1]{\boldsymbol{#1}}

\newcommand{\avec}{\vec{a}}
\newcommand{\ivec}{\vec{i}}
\newcommand{\jvec}{\vec{j}}
\newcommand{\kvec}{\vec{k}}

\newcommand{\coord}[2]{(#1, #2)}
\newcommand{\iina}{\ivec \,\, \mathrm{in} \,\, \mathcal{A}}

\newcommand{\erf}{\textrm{erf}}


\begin{document}

\title{Principled point-source detection in collections of astronomical images}
\author{Dustin Lang}
%\altaffiliation{To whom correspondence should be addressed; \texttt{dstndstn@gmail.com}}
% \affiliation{%
%   Dunlap Institute and Department for Astronomy \& Astrophysics,
%   University of Toronto,
%   50 Saint George Street, Toronto, ON, M5S 3H4, Canada}
\affiliation{%
  Perimeter Institute for Theoretical Physics,
  31 Caroline Street North, Waterloo, ON N2L 2Y5, Canada}
\affiliation{%
  Department of Physics \& Astronomy,
  University of Waterloo,
  200 University Avenue West, Waterloo, ON N2L 3G1, Canada}
\author{David W. Hogg}
\affiliation{%
  Center for Computational Astrophysics,
  Flatiron Institute,
  162 Fifth Ave, New York, NY 10010, USA}
\affiliation{%
  Center for Cosmology and Particle Physics,
  Department of Physics,
  New York University, 726 Broadway, New York, NY 10003, USA}
\affiliation{%
  Max-Planck-Institut f\"ur Astronomie,
  K\"onigstuhl 17, D-69117 Heidelberg, Germany}
\date{Dec 31, 2020}
\shorttitle{Point-source detection}
\shortauthors{Lang and Hogg}
%\correspondingauthor{Dustin Lang}
%\email{dstndstn@gmail.com}

\begin{abstract}
We review the well-known \emph{matched filter} method for the
detection of point sources in astronomical images.  This is shown to
be optimal (that is, to saturate the Cram\'er--Rao bound) under stated
conditions that are very strong: an isolated source in
background-dominated imaging with perfectly known background level,
point-spread function, and noise models.  We show that the matched
filter produces a maximum-likelihood estimate of the brightness of a
purported point source, and this leads to a simple way to combine
multiple images---taken through the same bandpass filter but with
different noise levels and point-spread functions---to produce an
optimal point source detection map.  We then extend the approach to
images taken through different bandpass filters, introducing the
\emph{SED-matched filter}, which allows us to combine images taken
through different filters, but requires us to specify the colors of
the objects we wish to detect.  We show that this approach is superior
to some methods traditionally employed, and that other traditional
methods can be seen as instances of SED-matched filtering with implied
(and often unreasonable) priors.
We present a Bayesian formulation, including a flux prior that leads to
a closed-form expression with low computational cost.
\end{abstract}

\section{Introduction}

There are few operations in astronomy more important than the
detection of stars or point sources.
Indeed, many astronomical discoveries come down to point-source
detection.
What is the best method for performing such detection?
Here we answer that question, in the limited context of isolated
sources, uniform sky-limited noise, and well-understood point-spread
function.
Even in this limited context, the subject is rich and valuable; more
general---and more difficult---cases will be illuminated if we can
understand the simplest case first.

Fundamentally, when much is understood about a signal latent in noisy
data, the best detection methods are (or look like) \emph{matched
filters}.
A matched filter is a model of the expected signal with which the data
are \emph{cross-correlated}. % (often wrongly called ``convolved'').
Peaks in the cross-correlation are candidate signal detections.
In point-source detection in astronomical images, the expected signal is the
point-spread function (PSF), and the cross-correlation operation is often
wrongly called ``convolution by the PSF''.
Matched filters are well used in astronomy, in contexts ranging from
spectroscopy \citep{bossspectro} to
galaxy clusters \citep{redmapper, melin} to
ultra-faint galaxies \citep{willman1} to
exoplanets \citep{exoplanet} to
gravitational radiation \citep{ligo}.

% from the tweeps:
% http://scholar.google.com/scholar?q=astronomy+matched+filter
% http://arxiv.org/abs/astro-ph/0602424

In what follows, we will argue for matched filtering for point-source
detection.  This is not new (it has been reviewed recently by
\cite{zackay1}); what is new is that we consider the common context of
heterogeneous (in point-spread function and sensitivity) multi-epoch,
multi-band imaging.
While the optimality of matched filtering for single-image point
source detection is well known by astronomers, the straightforward
mathematics behind it is often not, leading to a misconception that it
is simply an algorithmic choice.  Here, we show that this method is
optimal in the technical sense that it saturates the Cram\'er--Rao
bound under stated assumptions.  The mathematics are straightforward
and the resulting procedure is simple and computationally inexpensive.

Perhaps more controversially, we go on to argue that when imaging in
multiple bands is available, one should again use a matched filter,
now matched to the spectral-energy distribution (SED) of the sources
to be detected.  This SED-matched filtering, as we will call it, makes
explicit the assumptions that are implicitly embedded in any method
that attempts to detect sources by combining imaging from multiple
bands.

In the Real World, astronomers never precisely know their point-spread
function, their noise model, their flat-field (or other calibration
parameters), nor the spectral-energy distributions of the sources of
greatest interest.
Also, often, the sources of interest aren't point sources or perhaps
vary with time.
In these cases, we advocate parameterizing ignorance, and operating
with the union of all possibly appropriate matched filters.
We will fully execute this idea here when it comes to spectral-energy
distributions, but there are natural extensions to deal with
point-spread function, noise-model, calibration, and time-domain
uncertainties.

% FIXME -- DISCUSS each of the things listed above in the Discussion (or omit)!

A ``traditional'' approach for detecting sources in multi-epoch
imaging is to co-add the images and then run a detection algorithm on
the resulting coadd.  When the images have different point-spread
functions or noise properties, this method results in needless loss of
sensitivity; producing a coadd effectively forces the use of a
\emph{mismatched filter} rather than a matched filter.  We will show
that the correct procedure involves creating a weighted co-addition of
matched-filtered (smoothed) images.

% Often, astronomers ``co-add'' their imaging to find faint
% objects.  Technically, this step is only justifiable if the bandpass
% and point-spread function are the same for all images, and the images
% ought to be weighted somehow according to sensitivity.  Co-addition is
% not necessary for source detection, of course; it is possible to
% combine low-significance source-detection information coming from many
% images and do as well or better than co-addition.  

% An amusing
% conclusion of this project is that in the principled limit, the
% combination of source-detection information looks very much like a
% (weighted) co-addition of (smoothed) data!

That is, in what follows, we will detect sources as above-threshold
pixels or regions in a weighted co-add of PSF-correlated input images.
We will call this object a ``detection map''.  This detection map is
the best thing to use for source detection.  Once sources are
detected, of course, the detection map should be put aside, and source
properties (positions, colors, and so on) ought to be measured
(inferred) from the raw pixels in the collection of input images via a
likelihood function.  That measurement and likelihood function is
beyond the scope of this paper, but the subject of a parallel research
program (eg, \cite{unwise-phot}).

\section{Our image model}

We consider idealized astronomical images such as those obtained from
a CCD in typical broadband optical imaging.
Specifically, we will make the following strong assumptions
(and later we will relax many of them):
\begin{itemize}
\item the noise (coming from such sources as the Poisson distribution
  of the sky background, dark current, and read noise) is zero-mean,
  Gaussian, pixelwise independent, and of known constant variance.  The
  zero-mean assumption can be seen as assuming perfect background
  subtraction (sky estimation);
\item the image is well sampled;
\item the (perhaps tiny) image contains \emph{only} one point
  source with an unknown flux and position;
\item the source is centered within a pixel; %(we will relax this
  % later);
\item the point-spread function is spatially constant (across the
  possibly small image patch of interest) and known perfectly;
\item the device is linear and the photometric calibration of the
  image is known perfectly; that is, that it is possible to map from
  image ``count'' units back to physical units or a photometric
  standard;
\item the image is perfectly astrometrically calibrated;
\item the image is not contaminated by cosmic rays, stray light, bad
  pixels, bad columns, electronic artifacts, or any other of the many
  defects in real images.
\end{itemize}

Throughout this \doctype\ we assume a ``pixel-correlated'' PSF; we
consider the point-spread function to \emph{include} the effects of
pixelization.  In well-sampled images, we think of the image as being
a continuous function which, after being correlated by the PSF, is
sampled at delta-function pixel locations.  There is no need to think
of pixels as ``little boxes''; they are simply samples of an
underlying smooth two-dimensional signal.

% FIXME -- mention that the pixel-sampled PSF is the thing we actually
% have access to in the images. (ie, if we use known stars to construct
% a PSF model)

With these strong assumptions, we can write down the probability
distribution of each pixel value, which allows us to prove the
optimality of the methods we present.
We will consider a discrete image made up of a square array of pixels
$\jvec$, each of which has value $I_{\jvec}$, where we
use $\jvec$ as a two-dimensional focal-plane position, measured in
integer pixel units.
If the image contains a single point source, centered on
the pixel at position $\kvec$ and with constant flux resulting in a
total number of counts $f$, then the image is
\begin{eqnarray}
  I_{\jvec} &=& f \, \psfat{\jvec - \kvec} + \noise_{\jvec} \quad ,
  \label{eqn:image}
\end{eqnarray}
where $\psfat{\jvec-\kvec}$ is the point-spread function evaluated at
offset $\jvec-\kvec$ and $\noise_{\jvec}$ is per-pixel noise drawn
from a zero-mean Gaussian with known, constant, per-pixel variance $\sigma^2$.
We can also write this as
\begin{eqnarray}
  I_{\jvec} &\drawnfrom& \gaussx{f \, \psfat{\jvec - \kvec}}{\sigma^2}
  \nonumber
  \quad ,
\end{eqnarray}
meaning that $I_{\jvec}$ is drawn from a Gaussian distribution with
mean $f\, \psf(\jvec - \kvec)$ and variance $\sigma^2$.

\section{Detecting a point source in a single image}
\label{sec:detection}

%\subsection{Optimal linear detection}

%Here we consider a point source $k$ in a single image, as in
%\eqnref{eqn:image}.

% the map $M_{\jvec}$ can
% be computed by correlating the image with its PSF:
% \begin{equation}
% M_{\jvec} = \sum_{\iina} \psi_{\ivec} \, S_{\ivec + \jvec} \quad ,
% \end{equation}
% where, as before, $\mathcal{A}$ is the support of the PSF and
% $\psfat{\ivec} = \psi(\ivec)$ is an image of the PSF evaluated at
% integer pixel positions $\ivec$.

The \emph{matched filtering} operation, also known as ``smoothing by
the PSF'' or ``correlating by the PSF''\footnote{
  Or, often, ``convolving by the PSF'', being slightly careless with terminology.}
can be written as
\begin{eqnarray}
  M_{\jvec} &=& \sum_{\iina} \psfat{\ivec} \, I_{\ivec + \jvec}
  \nonumber \quad ,
\end{eqnarray}
where $\mathcal{A}$ is the support of the PSF and $\psfat{\ivec} =
\psf(\ivec)$ is an image of the PSF model evaluated at integer pixel
offset $\ivec$, where the PSF is centered at the origin.  This
operation can be seen as ``gathering up'' the signal that is dispersed
into many pixels by the PSF, weighting by the fraction of the flux
that went into the pixel.  In \appref{app:lindet} we derive the
matched filter and show that it saturates the Cram\'er--Rao bound.

% In practice, it is common to use a symmetric (circular) Gaussian
% approximation of the PSF model.  This has the computational advantage
% that one can use a fast (separable) real-space filtering routine, and 

% In \appref{app:lindet}\ we show that \emph{correlating} an image by
% its PSF---\emph{matched filtering}---results in an optimal detector
% for isolated point sources.  This operation can be seen as ``gathering
% up'' the signal that is dispersed into many pixels by the PSF; or as
% re-scaling each pixel in the support of the PSF into an estimate of
% the total flux and producing a weighted average of these estimates.

We define the \emph{\detmap} $D_{\jvec}$ as the matched filter, scaled to be in convenient units:
\begin{equation}
D_{\jvec} = \frac{1}{\psfnorm^2} \sum_{\iina} \psfat{\ivec} \,
I_{\ivec + \jvec} \quad ,
\label{eq:detmap}
\end{equation}
where the summation operation is the correlation of image $I$ by its
PSF $\psf$.
%where $\mathcal{A}$ is the support of the PSF and
%$\psfat{\ivec} = \psf(\ivec)$ is an image of the PSF model evaluated at
%integer-offset pixel positions $\ivec$.
The PSF norm $\psfnorm$ is defined as
\begin{equation}
\psfnorm = \sqrt{\sum_{\iina} \psfat{\ivec}^2} \quad ,
\end{equation}
and as shown in \appref{app:gaussnorm}, a Gaussian PSF with standard
devation $\psfw$ pixels has a norm approximately:
\begin{equation}
  \norm{\bm{\psf^G}} \simeq \frac{1}{2 \sqrt{\pi} \psfw} \quad .
\end{equation}
The per-pixel error in the \detmap\ is given by
\begin{equation}
\sigma_{D} = \frac{\sigma}{\psfnorm} \quad .
\end{equation}

We have scaled the \detmap\ so that each pixel contains the
maximum-likelihood estimate of the total flux of a source centered
at that pixel.  That is, if we compute at pixel $\jvec$ the flux
$f^{\ast}_{\jvec}$ that minimizes the chi-squared
($\chi^2$) residual within the support of the PSF:
\begin{equation}
  f^{\ast}_{\jvec} = \arg\min_{f} \sum_{\ivec} \left( \frac{I_{\ivec+\jvec} - f \, \psfat{\ivec}}{\sigma} \right)^2
  \nonumber
\end{equation}
we find
\begin{eqnarray}
  f^{\ast}_{\jvec} &=& \frac{\sum_{\ivec} I_{\ivec+\jvec} \, \psfat{\ivec}}{\sum_{\ivec} \psfat{\ivec}^2}
  \nonumber
  \\
  f^{\ast}_{\jvec} &=& D_{\jvec} %\quad .
  \nonumber
\end{eqnarray}
as defined above.
That is, a significant peak in the \detmap\ indicates the likely
position of a point source, and the value of the \detmap\ at a pixel
is the maximum-likelihood estimate of the flux for a source centered
at that pixel.

\subsection{Threshold and Peaks}

% In astronomical imaging, and in particular in large-scale surveys, it

Once we have computed a detection map, we typically wish to produce a
list of detected sources.  Standard practice is to apply a threshold
at, say, $5 \sigma_D$, and accept any peak above that threshold as a
source (or a blended group of sources).\footnote{%
  ``Deblending'' nearby groups of sources is a challenging task that
  is beyond the scope of this paper.}
In regions containing no
sources, the detection map contains Gaussian noise.  Due to the
correlation operation, the detection map pixels are not pixelwise
independent, but as weighted sums of Gaussian samples they are still distributed
as Gaussians.  As such, the expected number of pixels above a threshold
$\tau$ is the integral of the high tail of the normal distribution.
For $\tau = 5 \sigma_D$, the fraction of pixels above threshold due to
noise is about $2.9\times10^{-7}$, which seems tiny except that we
will be evaluating millions of pixels; in a 4k$\times$4k image we
would expect approximately $5$ false positive detections.

There is nothing special about $5 \sigma_D$ as a detection threshold; it
is simply a choice of tradeoff between allowing some false positives
while preventing too many false negatives (lost detections).  In
different situations, higher or lower thresholds could be preferable.

\subsection{Comments}

\paragraph{PSF model}
Computing the detection map requires correlation of the image by a
model of its point-spread function.  In practice, the PSF model is
never known exactly, and since correlation by large pixelized models
can be expensive, it is common to approximate the PSF by a Gaussian
for the purposes of detecting sources.  The impact of this
approximation on detection efficiency is apparent in the derivation of
the matched filter; in equation \ref{eqn:psfdotprod}, the detection
map signal-to-noise is proportional to the cosine distance between the
true PSF and the correlation kernel.  In typical ground-based images,
this results in only a few percent loss in signal-to-noise.  For
example, in our DESI Legacy Imagine Surveys images \citep{lsoverview},
if we assume that our pixelized
model of the PSF is correct, then a Gaussian approximation typically
yields above $97\%$ efficiency.  A considerable side benefit of making
this assumption is that one can use a fast separable real-space
filtering routine to perform the correlation operation.

\paragraph{Biases}
In this paper, we have assumed that backgrounds due to atmospheric
emission (``sky'') and detector effects such as bias and dark current
have been perfectly estimated and subtracted.  In real images,
however, errors in these estimates can leave spatially coherent
residual biases, and these can have a considerable effect on source
detection.  For example, a background that is elevated by $0.05
\sigma$ per pixel can double the false positive rate (with a
$5\sigma_D$ threshold) in good seeing, and has an even greater effect
in worse seeing.

\paragraph{Sub-pixel peaks}  The detection map defined above
is computed by correlating the image with its PSF, on the image pixel
grid.  As the PSF becomes narrow, the detection efficiency varies
depending on the position of a source within the pixel.  The detection
map is maximized when the image is a scaled version of the PSF
(matched filter), but if the image is shifted within a pixel relative
to the PSF model, then the peak value attained by the detection map is
lower (because the filter is slightly mismatched).  For example, with
a Gaussian PSF with standard deviation of 1 pixel, the detection map
drops as low as 88\% efficiency for a source midway between pixels in
both dimensions.

To reduce this effect, one could compute multiple detection maps,
using for each a different subpixel-shifted versions of the PSF model.
Alternatively, one could lower the detection threshold to compensate,
then fit for the best source position and drop sources with best-fit
values below threshold.

\paragraph{Sufficient statistic}  Correlation by the PSF summarizes
all relevant information regarding the presence of a source at each
pixel; the detection map and its variance are sufficient to describe
our knowledge.  In some astronomical source detection packages
(including SourceExtractor), there is a notion of requiring more than
one neighboring pixel to exceed a detection threshold.  This is not
necessary or useful; in effect it imposes a larger detection threshold
that varies based on the source morphology and PSF, which is
typically undesirable.

% \paragraph{What would Bayes do?}  FIXME -- We haven't really explained that
% we are producing and evaluating an estimate for each pixel.  Could
% also point out that one could instantiate and test the generative
% model (with position and flux) at each pixel, and that this basically
% just short-cuts that process, telling you places where that test is
% going to give significant results.

\paragraph{Galaxies}
We have focused only on point sources, but the same arguments can be
used to develop a detector for galaxies.  The matched filter by which
the image must be correlated is then the intrinsic galaxy profile
correlated with the PSF.  It turns out \citep{zackay1} that a
matched-filtered image for a given galaxy profile can be computed by
correlation with the PSF detection map.
Of course, a matched-filtering approach is only optimal
for a single galaxy profile.  As with using an approximation for the
PSF model, the non-optimality due to using an incorrect galaxy profile
is related to the cosine distance between the true and model galaxy
profiles.

In practice, our mixture-of-Gaussians approximations to standard
exponential and deVaucouleurs galaxy profiles \citep{gaussgals} are
convenient for this task.

\section{Detecting a point source in multiple images}

In this section we will assume we have a stationary point source whose
flux is constant over time, and a series of images taken through
different bandpass filters and with different noise levels, exposure
times, point-spread functions, and telescope pointings.  We can
achieve optimal detection of the source by building a \detmap\ for
each image and combining them with weights as described below.

\subsection{Identical bandpass filters}

We first present the simpler case where all the images are taken
through identical bandpass filters.

As we have seen, the \detmap\ defined in \eqnref{eq:detmap} is a
maximum-likelihood estimate of the total \emph{counts} contributed by
the source, in the units of the original image.  In order to
combine information from multiple images, we must calibrate them so
that they are in the same units.  Since this calibration is simply a
linear scaling, it can be applied to the original image or to the
\detmap.  Similarly, if the images are on different pixel
grids---either from different pointings of the same CCD, or from
different CCDs---then we must \emph{resample} the \detmap s to a
common pixel grid.
If the original image is well-sampled, then the \detmap\ (which has
been further smoothed by PSF correlation) will also be well-sampled,
so resampling to a pixel grid of the same or finer resolution results
in no loss of information.
Since the pixel values in the \detmap\ represent the \emph{total} flux
from a point source, the \detmap\ does not need to be rescaled when
resampled to a different pixel scale.

Once the \detmap\ for each image has been calibrated and resampled to
a common pixel grid, we have multiple \emph{independent}
maximum-likelihood estimates of the source flux in our chosen filter,
each with a known standard deviation and Gaussian statistics.  That
is, we have multiple Gaussian likelihood functions that we wish to
combine.  Since they are independent, the combined likelihood is the
product of the individual likelihoods.  For Gaussian distributions,
the resulting aggregate maximum likelihood estimate is the
inverse-variance-weighted sum of the individual estimates.

If the calibration factor $\kappa_i$ scales image $i$ to flux in
common units, and $R_i$ represents resampling to the common pixel grid,
then the flux estimate $F_i$ is
\begin{eqnarray}
F_i &=& R_i(\kappa_i \, D_i)
\end{eqnarray}
with per-pixel error
\begin{eqnarray}
\sigma_{F_i} &=& \frac{\kappa_i \, \sigma_{i}}{\psfnorm_i}
\end{eqnarray}
and we combine the estimates from multiple images via
\begin{eqnarray}
F^{\star} &=& \frac{\displaystyle\sum_i F_i \, \sigma^{-2}_{F_i}}{\displaystyle\sum_i \sigma^{-2}_{F_i}}
\label{eq:onebandmap}
\end{eqnarray}
which has per-pixel error
\begin{eqnarray}
  \sigma_{F_{\star}} &=& \left( \sum_i \sigma^{-2}_{F_i} \right)^{-\frac{1}{2}}    \quad .
  \label{eq:onebandstd}
\end{eqnarray}
This is simply the maximum-likelihood estimate of the flux based on a
set of independent Gaussian estimates.

In summary, the procedure to produce an optimal detection map given
multiple images (in the same filter) is:
\begin{enumerate}
\item \emph{correlate} each image by its PSF model
\item \emph{calibrate} each resulting detection map (and its variance)
  to common units
\item \emph{resample} each calibrated detection map to a common pixel
  grid
\item \emph{coadd} the calibrated detection maps weighted by their
  inverse variances.
\end{enumerate}
Assuming well-sampled images, the \emph{correlation},
\emph{calibration}, and \emph{resampling} steps can occur in any
order.  Importantly, however, the \emph{coaddition} stage must occur
\emph{after} correlation by the point-spread functions of the
individual images; each image must be correlated by its own matched
filter to produce \emph{detection maps} which are then coadded.

\subsection{Comments}

\paragraph{Optimality}
In appendix \ref{app:multiopt} we show that the estimator $F^{\star}$
saturates the Cram\'er--Rao bound and is therefore statistically
optimal.

\paragraph{Coadds}
Occasionally, astronomers attempt to construct image coadds and then
detect sources by correlating the coadd with an estimate of its PSF.
It is straightforward to show that this is necessarily sub-optimal
unless the images have the same PSF.  Intuitively, the PSF of the
coadd is not equal to the PSF of either image, therefore this approach
uses a ``mismatched filter'' rather than a matched filter, resulting
in loss of signal-to-noise or detection efficiency.

\begin{figure}[h]
  \begin{center}
    % This figure comes from the dont-coadd.py script.
  \includegraphics[width=0.5\textwidth]{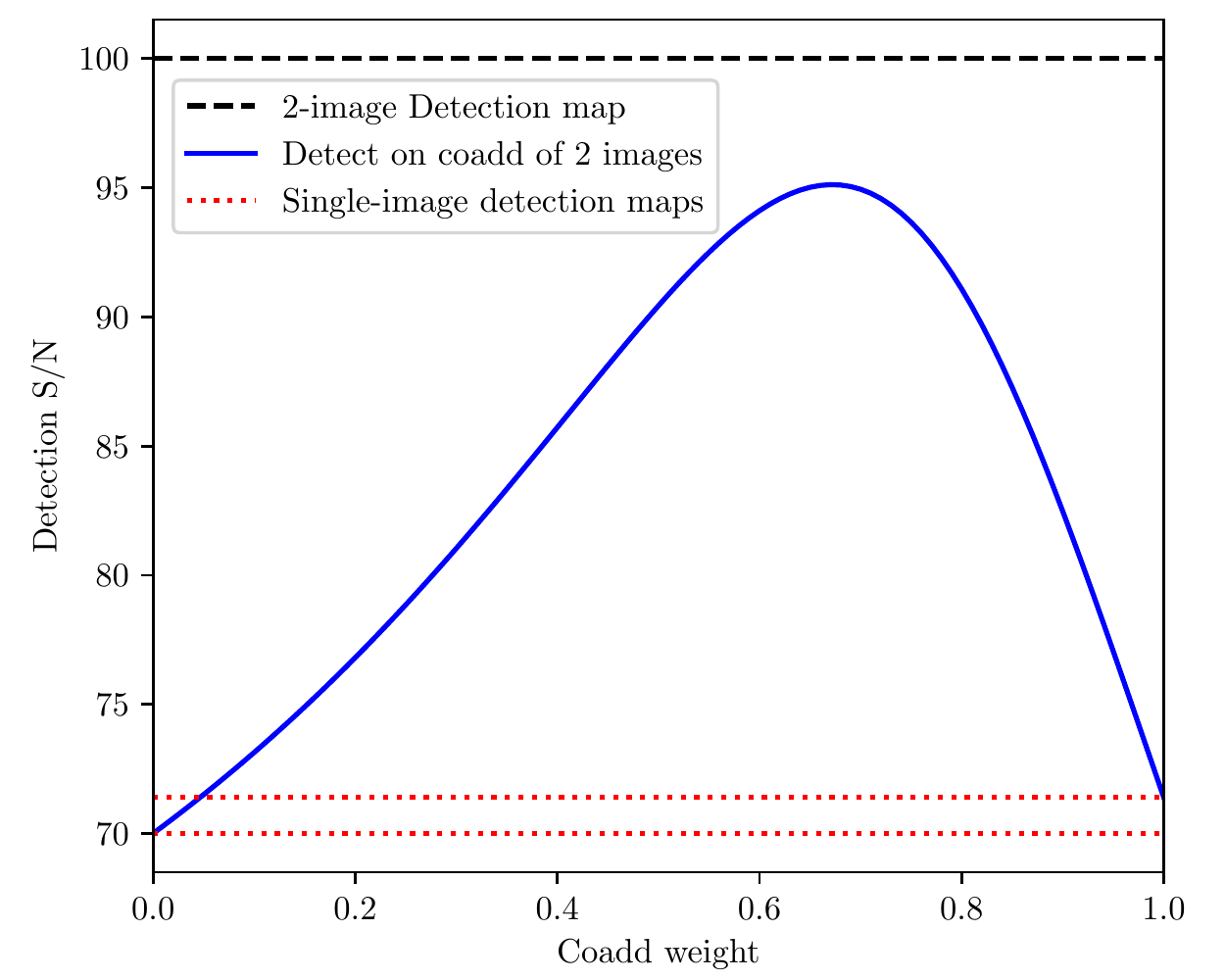}
  \caption{The benefit of constructing a \emph{detection map} versus
    \emph{coadding} the images and then detecting sources on the
    coadd.  Here, we have two images with PSF widths different by a
    factor of two, and exposure times such that the depths are
    similar.  The signal-to-noise at which a source is detected in the
    individual images is shown by the dotted lines at the bottom.  By
    constructing a detection map, we extract all the available signal
    in the combination of the two images; the detection map
    signal-to-noise (dashed line at top) equals the sum-in-quadrature
    of the two images.  However, if we instead coadd the images and
    then detect the source in the coadd, we lose a significant
    fraction of the signal-to-noise, regardless of the weighting
    factor applied to the two images.
    \label{fig:dontcoadd}}
  \end{center}
\end{figure}

As an illustration, we simulated two images with similar detection
signal-to-noise but Gaussian PSFs that differed by a factor of two.
We computed the detection map, and also created a series of coadds
(trying different weights for the two images), computing the detection
map for each, using the correct coadded PSF.  As shown in Figure
\ref{fig:dontcoadd}, regardless of the coadd weight chosen, detecting
on the coadd results in a loss of efficiency compared to the detection
map.

\subsection{Different bandpass filters: the \emph{SED-matched filter}}
\label{sec:sedmatched}

Everything we have said up to now has been based on facts about
statistical distributions and should be uncontroversial.  In this
section we propose a method that is, to our knowledge, new to
astronomy, though it flows naturally from the multi-image matched
filtering we have discussed. While it is fully defensible, it involves
\emph{priors} so we expect will be slightly controversial.  We argue
that other proposed methods presume \emph{stronger} and usually
\emph{unstated} priors.

As we saw in the single-bandpass case, we can combine multiple
individual exposures into an aggregate estimate of the flux of a point
source.  In order to do this, it was essential to calibrate the images
so that each one was an estimate of the same underlying quantity.  The
multiple-bandpass case is similar: For each bandpass, we first combine
the images taken in that bandpass into an aggregate estimate.  Then,
to combine the bandpasses we must scale them so that they are
estimates of the same quantity.  This requires \emph{knowing} the
spectral energy distribution, or at least the \emph{colors} in the
filters of interest, of the source to be detected; this allows us to
scale the various bandpasses so that they are estimates of a common
quantity: perhaps the flux in a canonical band, or some other linear
quantity such as the integrated intensity.

%% FIXME -- does that make sense?  Apparent brightness?  Apparent
%% luminosity?  I just want to say, you could measure it in anything
%% linear, like W/m^2/sr/s or, heck, the projected area of a star of a
%% given temperature)

The intuition here is that if we know that our sources of interest are
twice as bright in bandpass A as in bandpass B, then we can convert an
estimate of the brightness in band B into an estimate of the
brightness in band A by multiplying by two.  The variance of the
scaled estimate increases appropriately (by a factor of four), so a
bandpass in which a source is expected to be faint will contribute an
estimate with a large variance and will be downweighted when the
estimates are combined.  We can also view the problem as one of
estimating a total flux that has been split into the different
bandpasses, and in that view the SED-matched filter is analogous to
the way flux is spread into pixels by the point-spread function (and
re-collected by correlating with the matched filter).

\newcommand{\sigdj}{\sigma_{D_j}}

Assume we have computed detection maps $D_j$, with per-pixel standard deviation
$\sigdj$, for a number of different bandpasses.  Assume each
bandpass has a known conversion factor $s_j$ to the canonical band;
that is,
\begin{eqnarray}
  D_j & \drawnfrom & \gaussx{F s_j}{\sigdj^2}
\end{eqnarray}
for flux $F$ in the canonical band.
% \footnote{% The notation $x
%   \drawnfrom \gaussx{\mu}{\sigma^2}$ means that $x$ is drawn from a
%   Gaussian distribution with mean $\mu$ and variance $\sigma^2$.}
%
Given a number of such detection maps, we first scale them so they are all estimates of the
same quantity, by dividing by $s_j$:
\begin{eqnarray}
  F_j & = & \frac{D_j}{s_j} \\
  F_j & \drawnfrom & \gaussx{\frac{D_j}{s_j}}{\frac{\sigdj^2}{s_j^2}}
\end{eqnarray}
and assuming that the images  are independent, we combine them by inverse-variance weighting
to produce the maximum-likelihood estimate for $F$:
\begin{eqnarray}
  \hat{F} &=& 
  %\frac{\sum_j D_j \, f_j \, \sigdj^{-2}}%
  %     {\sum_j f_j^2 \, \sigdj^{-2}}
  \frac{\displaystyle\sum_j \frac{D_j}{s_j} \, \frac{s_j^2}{\sigdj^2}}%
       {\displaystyle\sum_j \frac{s_j^2}{\sigdj^2}}
       = 
       \frac{\displaystyle\sum_j D_j \, s_j \, \sigdj^{-2}}%
            {\displaystyle\sum_j s_j^2 \, \sigdj^{-2}}
\end{eqnarray}
with per-pixel error
\begin{eqnarray}
  \hat{\sigma}_F &=& \left( \sum_j s_j^2 \sigdj^{-2} \right)^{-\frac{1}{2}}
  \quad .
\end{eqnarray}

For example, if we treat $r$ band as the canonical band and our
objects of interest have color $r-i = 1$, then we expect the flux in
$i$ to be a factor of $2.5$ greater than the flux in $r$; $s_i = 2.5$,
and we will scale our $i$-band detection map $D_i$ by $1/s_i = 0.4$ to
produce a prediction for the $r$-band flux.  Since the sources are
expected to be brighter in $i$ band, we must \emph{scale down} the
$i$-band estimate to produce an $r$-band estimate.  The $i$-band
variance is also reduced in a corresponding way, so this does not
dilute the weight of high-precision measurements.

In practice, for those unwilling to use the Bayesian method presented
below, we would advocate computing SED-matched detection maps for a
set of spectral energy distributions that sparsely sample the space of
sources of interest, and take the union of sources detected.  By
performing multiple significance tests, more false positives will be
generated, so it will be necessary to slightly increasing the
detection threshold to maintain false positives at an acceptable
level.  Since we effectively extract more of the available
signal-to-noise by weighting the bands appropriately, our method
should still achieve superior completeness at a given purity.

More broadly, we argue that source detection should be used as in
\emph{initial} seed of the likely positions of sources, but that only
after \emph{inference} of the proposed source's properties using the
individual images should one decide whether the source should be kept.
This leads toward using slightly lower detection thresholds (that will
produce more false positives due to noise), plus additional
thresholding after fitting to determine which sources should be kept.

\subsection{Comments}

\begin{figure}
  \begin{center}
    % This figure comes from detection-figs.py : chisq_fig()
    \includegraphics[width=0.8\textwidth]{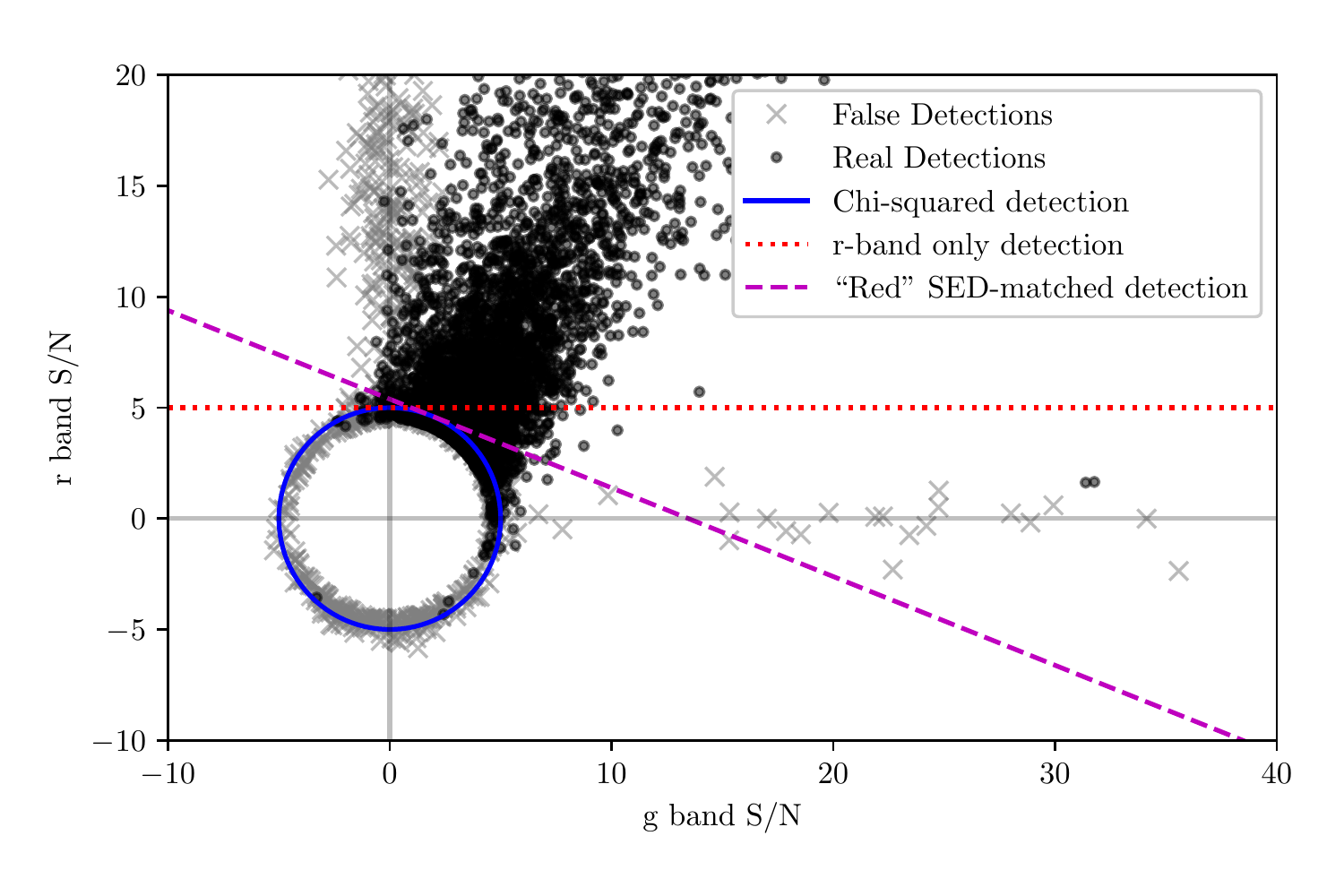}
    \caption{SED-matched versus ``chi-squared'' detection filters.
      The data points are (chi-squared) detections above $4.5 \sigma$
      in one $g$-band and one $r$-band image taken with the Dark
      Energy Camera, plotted in signal-to-noise space.  Points that
      are not detected in deeper data are marked
      as ``False''; these are mostly clustered around zero (due to
      noise), with some scattered points near the axes due to
      single-band artifacts such as cosmic rays. The main locus of
      ``Real'' peaks correspond to real stars and galaxies with
      typical colors.  The Chi-squared detection method selects all
      sources outside the circle (including, in the naive formulation,
      sources with negative flux in both bands).  A single-band
      detection filter selects all sources above a vertical or
      horizontal detection threshold.  Our ``Red'' SED-matched filter
      selects sources where the fluxes in the two bands are consistent
      with a red SED (in this case, a $g - r$ color of 1).  Note that
      the threshold line is roughly orthogonal to the line between the
      false positives and the true positives; it separates them
      efficiently.  We have shown the chi-squared detection filter
      with a threshold of $(5\sigma)^2$, which yields a higher false
      positive rate than a $5\sigma$ threshold in a single-band or
      SED-match filter.
      \label{fig:sedmatched}
    }
  \end{center}
\end{figure}

\paragraph{Chi-squared coadd.}
\cite{szalay1999} present the idea of using the $\chi^2$
statistic of a set of images taken through different bandpass filters.
That is, they take pixel-aligned and possibly PSF-filtered images
(here we will use detection maps) and compute $\chi^2$ pixelwise
over bands $j$,
\begin{eqnarray}
  \chi^2 &=& \sum_j \frac{D_j^2}{\sigma_{D_j}^2}
\end{eqnarray}
and a sufficiently large set of connected pixels with $\chi^2$ above a
detection threshold is taken as evidence as a source.  See Figure
\ref{fig:sedmatched}.

As \cite{szalay1999} state, the $\chi^2$ detection method represents
the probability that a pixel is drawn from the Gaussian sky background
distribution (independently in each band).  A large value is
considered to reject this null hypothesis.  The ``source'' hypothesis
is not stated, but implicitly, by choosing a constant $\chi^2$
threshold, a source is assumed to have any non-zero flux uniformly
distributed on the surface of the signal-to-noise (hyper-)sphere.  As
noted in the paper, this includes sources with negative fluxes in
every band.  While they suggest heuristics to trim such sources (for
example, demanding $> -1 \sigma$ of flux in each band), the fact that
these sources are detected in the first place hints at the primary
issue with this method: that the ``source'' hypothesis is not
physical.

Lacking a source model means that it is not always helpful to add more
data: Consider the case where we have one informative band and several
noisy (but still somewhat informative) bands.  The chi-squared method
treats all bands equally, thus mixes the one informative band with all
the uninformative bands.  When using multiple bands, the detection
threshold must be increased to maintain a constant false detection
rate (as detailed below), and therefore the number of true detected
sources will be lower.

As shown in Figure \ref{fig:sedmatched}, there is one additional
caveat for the chi-squared detection method: a $(5\sigma)^2$ threshold
yields a larger false positive rate than a standard single-band
detection filter with a threshold of $5\sigma$; in order to achieve a
desired false positive rate, the detection threshold must be set by
analysis (with two bands, the survival function of the chi
distribution at the equivalent of $5 \sigma$ for a Gaussian is roughly
5.5; with three bands it is roughly 5.75) or simulations (as suggested
by \cite{szalay1999}).

\cite{szalay1999} in fact also suggest a method for detecting objects
of a specific color that is similar but not identical to the approach
we present here; they suggest projecting the multiple bands into
subspaces and using their chi-squared approach in those subspaces,
which means that the issues identified above still hold.

\subsection{Going Bayesian}

The \emph{SED-matched filter} presented above tells us how to find
likely source positions given a source spectral energy distribution.
It is then natural (in a Bayesian framework) to \emph{marginalize}
over the SED using a prior distribution.  We can also marginalize over
the flux of the source, allowing us to compare the hypothesis that a
source exists, versus the hypothesis that the observed flux is due to
a noise fluctuation.

We can write the likelihood for a single pixel in the set of detection
maps $D_j$ for bands $j$, given the existence of a source, as
\begin{equation}
  p_S(\{ D_j \}) = \iint p(\{ D_j \} | F, s) \, p(F) \, p(s) \, \dd F \, \dd s
\end{equation}
where $F$ is the flux of the source in some canonical band, and $s$ is
the SED of the source; here, we have made the strong assumption that the
flux and SED priors are independent.  We will assume that the SED prior $p(s)$ is
represented as a weighted sum of discrete SEDs: a gridding of SED
space, for example.  We then have
\begin{eqnarray}
  p_S(\{ D_j \})
  &=&
  \sum_{i} w_i \int p(\{ D_j \} | F, s_i) \, p(F) \, \dd F
  \\
  &=&
  \sum_i w_i \int \prod_j \gaussian{D_j \,|\, F \cdot s_{i,j}, \sigma_{D_j}^2} \, p(F) \, \dd F
  \\
  &=&
  \sum_i w_i 
  \left( \prod_j \frac{1}{\sqrt{2 \pi} \sigma_{D_j}^2} \right)
  \int 
    \exp{ \left( \sum_j \frac{(D_j - F \cdot s_{i,j})^2}{-2 \, \sigma_{D_j}^2} \right) }
    \, p(F) \, \dd F
\end{eqnarray}
where we have written out the indepedent Gaussian likelihoods of the
detection maps\footnote{The notation $\gaussian{x \,|\, \mu, \sigma^2}$ indicates the likelihood
  of drawing value $x$ from the Gaussian distribution $\gaussian{\mu, \sigma^2}$.},
and the SEDs are represented as scalings of the
canonical flux $F$: for SED $i$, band $j$ is predicted to have flux $F
\cdot s_{i,j}$.

We must now specify a prior over the flux to make progress.  One
option that leads to a closed-form result is an exponential prior,
$p(F) = \alpha \exp(-\alpha F)$ and $F > 0$, with $\alpha$ a free
variable (to be chosen).  We caution that this prior does have some
undesirable properties, discussed below.  With this exponential flux
prior, we have
\begin{eqnarray}
  p_S(\{ D_j \})
  &=&
  K \alpha
  \sum_i w_i 
  %\left( \prod_j \frac{1}{\sqrt{2 \pi} \sigma_j^2} \right)
  %\exp{\left( \sum_j -\frac{d_j^2}{2 \sigma_j^2} \right)}
  %\, K %k \, \chi^2 \,
  \int 
    \exp{ \left( F \sum_j \frac{D_j s_{i,j}}{\sigma_{D_j}^2} \right)}
    \exp{ \left( -F \alpha \right)}
    \exp{ \left( F^2 \sum_j \frac{s_{i,j}^2}{\sigma_{D_j}^2} \right)}
    \, \dd F
\end{eqnarray}
%where we have introduced the constants
%\begin{eqnarray}
%k &=& \prod_j \frac{1}{\sqrt{2 \pi} \sigma_j^2}
%\\
%\chi^2 &=& \exp{\left( \sum_j -\frac{d_j^2}{2 \sigma_j^2} \right)}
%\end{eqnarray}
where we have pulled out the constant
\begin{eqnarray}
K &=& \prod_j \frac{1}{\sqrt{2 \pi} \sigma_{D_j}^2} 
\exp{\left( \sum_j -\frac{D_j^2}{2 \sigma_{D_j}^2} \right)}
%\\
%&=& \gaussian{D_j | 0, \sigma_j^2}
= \gaussian{D_j \,|\, 0, \sigma_{D_j}^2}
\label{eq:pbg}
\end{eqnarray}
which will be recognized as the zero-mean Gaussian probability of
$\{D_j\}$: the likelihood that data values $D_j$ are drawn from the
background distribution!

Defining variables
\begin{eqnarray}
  a_i &=& \alpha - \sum_j \frac{D_j s_{i,j}}{\sigma_{D_j}^2} \\
  b_i &=& \frac{1}{2} \sum_j \frac{s_{i,j}^2}{\sigma_{D_j}^2}
\end{eqnarray}
we get an integral in which the flux prior can be integrated analytically:
% eg
% http://www.wolframalpha.com/input/?i=integrate+e%5E-ax+e%5E-bx%5E2+dx+from+x%3D0+to+%2Binfinity
% integral_0^(+∞) e^(-a x) e^(-b x^2) dx =
% (sqrt(π) e^(a^2/(4 b)) erfc(a/(2 sqrt(b))))/(2 sqrt(b))
% IFF b>0 OR (b>=0 AND a>0)
% (phew, b>0 always)
%
\begin{eqnarray}
  p_S(\{ D_j \})
  &=&
  K \alpha
  \sum_i w_i 
  \int_0^{\infty}
    \exp(-a_i F) \exp(-b_i F^2)
    \, \dd F
    \\
  p_S(\{ D_j \})
  &=&
  K \alpha
  \sum_i w_i 
  \frac{\sqrt{\pi}}{2 \sqrt{b_i}}
  \exp \left(\frac{a_i^2}{4 b_i} \right)
  \left(1 - \erf\left( \frac{a_i}{2 \sqrt{b_i}} \right) \right)
  \label{eq:pfg}
\end{eqnarray}
which can be evaluated numerically with modest computational cost as
long as the number of SEDs $i$ is not too large.  In practice, a
coarse gridding of SED space yields good results, as shown below.

% and defining variables
% \begin{eqnarray}
%   \beta_i &=& 2 \sqrt{b_i} = \sqrt{2 \sum_j \frac{s_{i,j}^2}{\sigma_j^2}}
%   \\
%   %c_i &=& \frac{a_i}{2 \sqrt{b_i}}
%   c_i &=& \frac{a_i}{\beta_i}
% \end{eqnarray}
% we can write
% \begin{eqnarray}
%   p_S(\{ d_j \})
%   &=&
%   K \alpha
%   \sum_i w_i 
%   \frac{\sqrt{\pi}}{\beta_i}
%   \exp(c_i^2) \left[ 1 - \erf(c_i) \right]
% \end{eqnarray}

% - a_i is negative for sources; b_i ~ constant across the image
% - a_i / (2 sqrt(b_i)) goes from slightly above zero to, eg, -50 for sources
% - erf(a_i / (2 sqrt(b_i))) ~ -1 for sources
% - 1 - erf(a_i / 2 sqrt(b_i)) ~ 2 for sources

In order to select sources, we can compare this likelihood to the
null-hypothesis likelihood: that there is no source and the observed
values are due to noise.  Conveniently, the null-hypothesis likelihood
is exactly the factor $K$ in the above expression!  Determining the
threshold at which to accept a peak in the $p_S/K$ map as a source can
be framed as a Bayesian decision theory problem, or can be tuned on
simulations to yield an acceptable false positive rate.

%% FIXME -- can we do it analytically?

\begin{figure}
  \begin{center}
    % These figures come from bayes-figure.py.
    \includegraphics[width=0.4\textwidth]{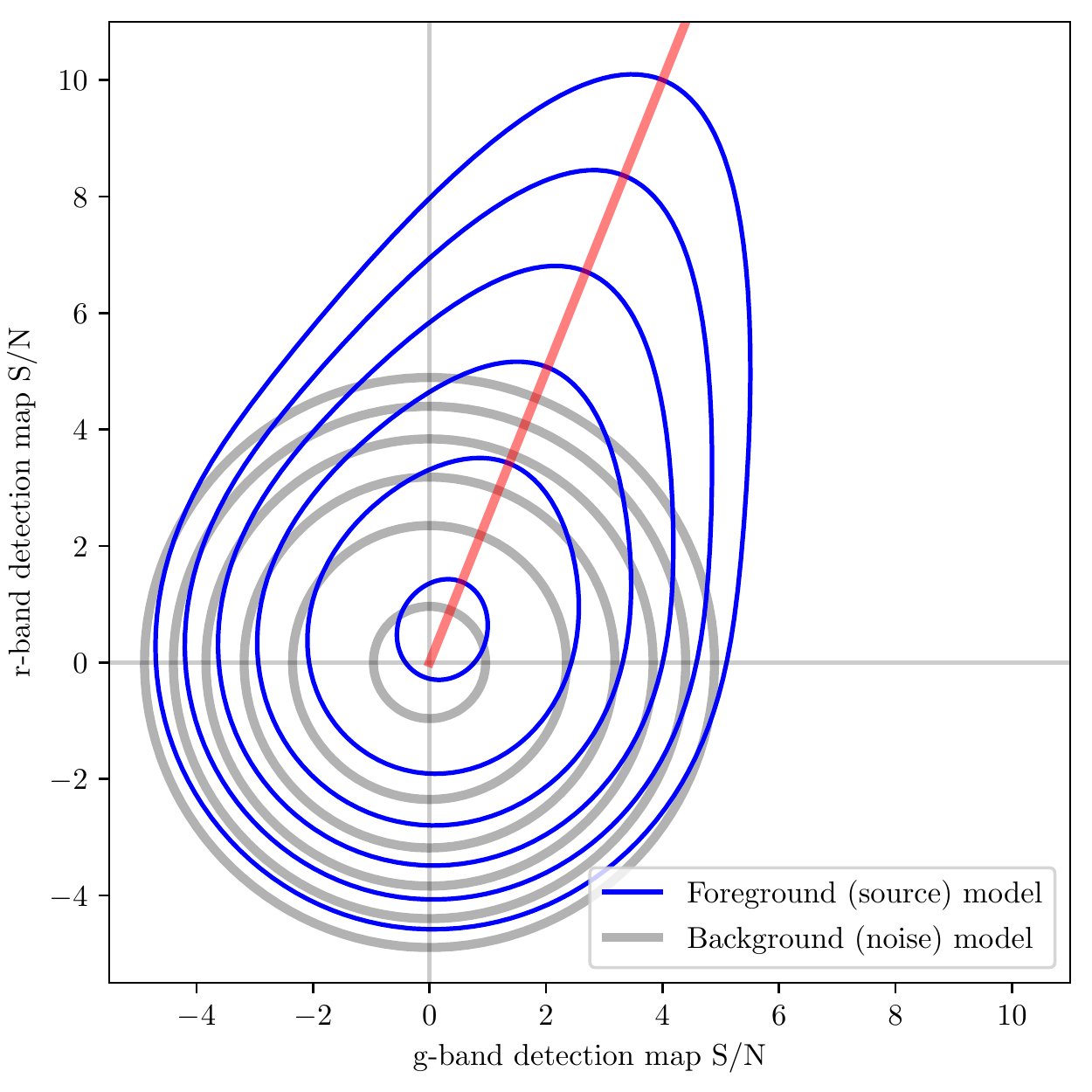}
    \includegraphics[width=0.4\textwidth]{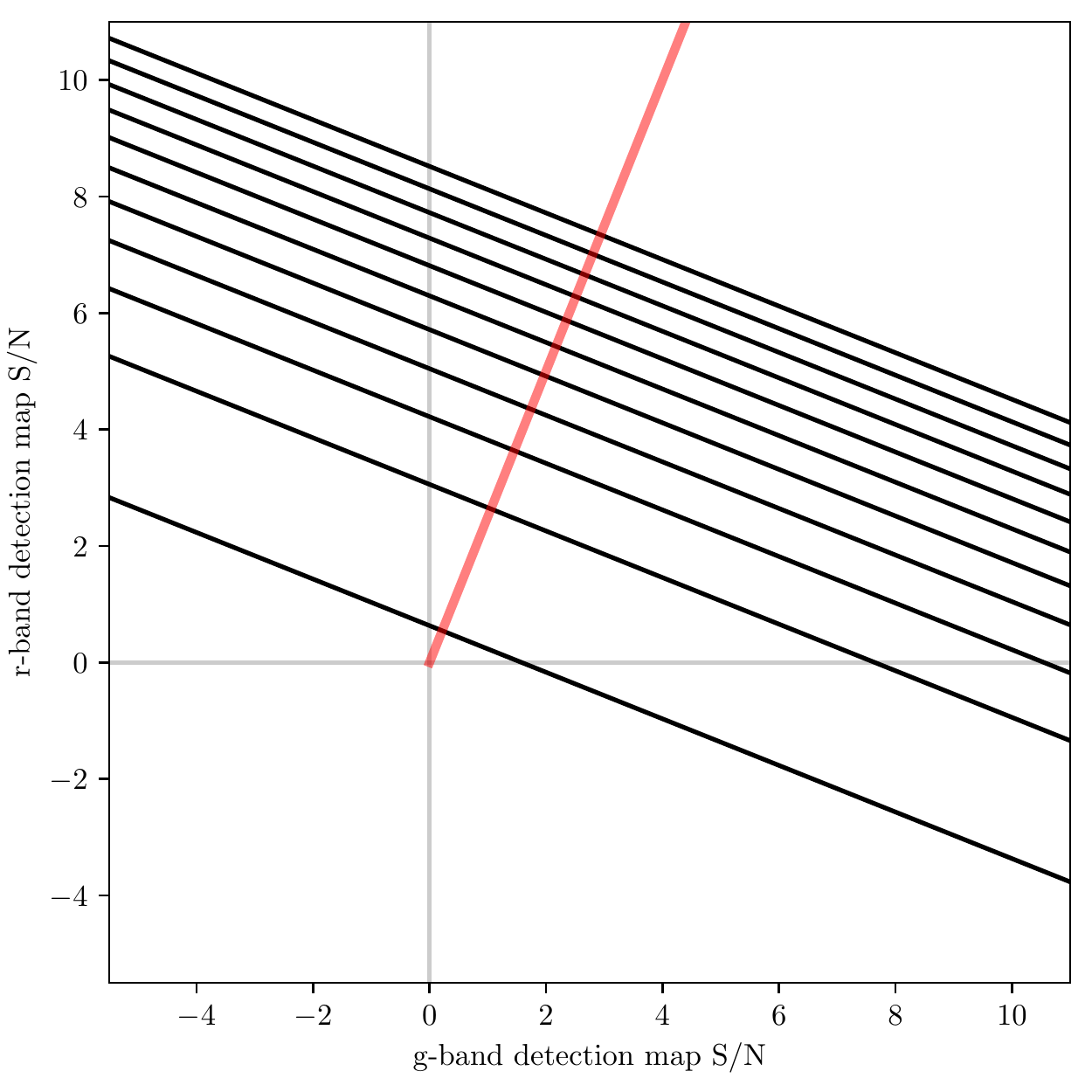}
    \caption{Bayesian SED-matched source detection with a single-SED prior, for illustration.
      The foreground model is that sources have fluxes that lie along a ``Red''
      SED with color $g-r = 1$.
      \emph{Left:} Probability contours for the foreground model are inclined in the
      direction of the prior.  The flux prior prefers small fluxes, but has some
      probability mass extended toward larger fluxes.
      The background model (null hypothesis) is that there is no source, only noise; these
      contours are circles centered on zero.
      The probability contours are spaced in powers of 10.
      \emph{Right:} Probability ratio contours for the
      foreground divided by background model.
      When performing source detection, these lines mark decision boundaries;
      points above and to the right of a decision boundary are
      considered to be detected.
      The contours are spaced in powers of 10.
      \label{fig:cona}
    }
    \end{center}
\end{figure}

\begin{figure}
    \begin{center}
    % These figures come from bayes-figure.py.
    \includegraphics[width=0.4\textwidth]{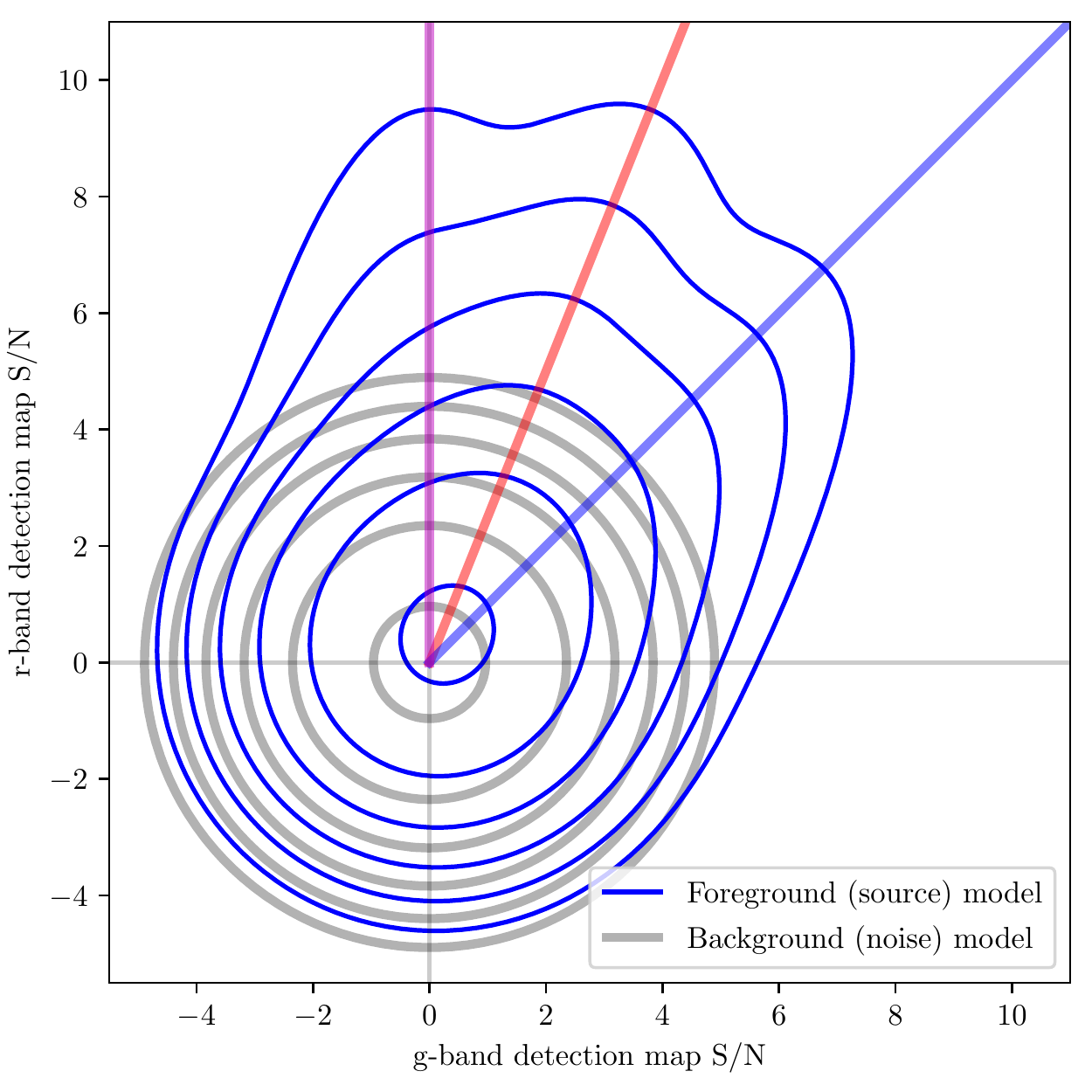}
    \includegraphics[width=0.4\textwidth]{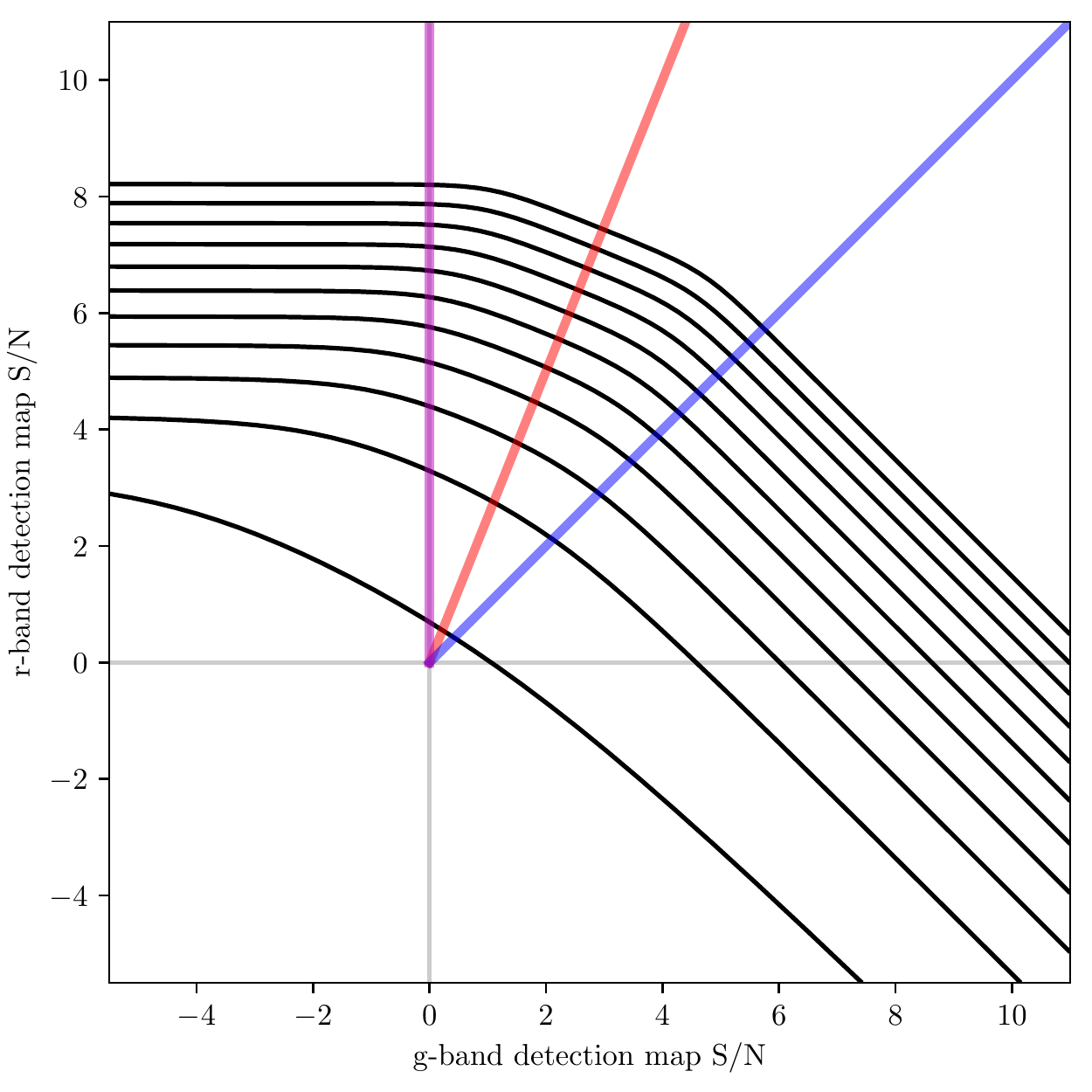}
    \caption{Bayesian SED-matched source detection with a three-SED prior.
      The prior is that astronomical sources are a
      mixture of $49\%$ ``Red'' (color $g-r = 1$ mag),
      $49\%$ ``Flat'' (color $g - r = 0$ mag),
      and $2\%$ $r$-band only ($g$ flux is zero).
      \emph{Left:} Probability contours for the foreground model are extended in the
      directions of the three components of the prior.
      The background model (null hypothesis) is that there is no source, only noise; these
      contours are circles centered on zero.
      The probability contours are spaced in powers of 10.
      \emph{Right:} Probability ratio contours for the
      foreground divided by background model.
      When performing source detection, these lines mark decision boundaries;
      points above and to the right of a decision boundary are
      considered to be detected.
      The contours are spaced in powers of 10.
      \label{fig:conb}
    }
    \end{center}
\end{figure}

To illustrate the method, we consider a case where we have two bands
of imaging, $g$ and $r$ bands.  In \figref{fig:cona}, we plot the
probability contours of the background model (equation \ref{eq:pbg})
and the foreground model (equation \ref{eq:pfg}), for the simplest
case where our SED prior $p(s)$ is a delta function at $s_{g} =
\frac{1}{3.5}$, $s_{r} = \frac{2.5}{3.5}$: this is a ``Red'' detection
filter that assumes a $g - r$ color of 1 magnitude.  As expected, the
probability contours of the foreground model shift away from negative
fluxes and toward fluxes that are consistent with the expected color.
The exponential prior on flux causes the foreground model to prefer
sources with small fluxes, and therefore the probability mass of the
foreground model is clustered around zero.  The probability ratio
contours---which are what would be used to accept or reject a proposed
source---are, as expected, orthogonal to the expected SED vector.

In \figref{fig:conb}, we show probability contours for a three-SED
model, including the ``Red'' SED and a ``Flat'' SED ($r$ flux equals
$g$ flux) with equal weights of $0.49$, plus an SED that has only flux
in the $r$ band, with weight $0.02$.  The probability contours of this
model are, as before, extended in the directions of these SEDs.  The
probability ratio contours transition smoothly between being
orthogonal to these three SED directions.

\begin{figure}
    \begin{center}
    % These figures come from bayes-figure.py.
    \includegraphics[width=0.45\textwidth]{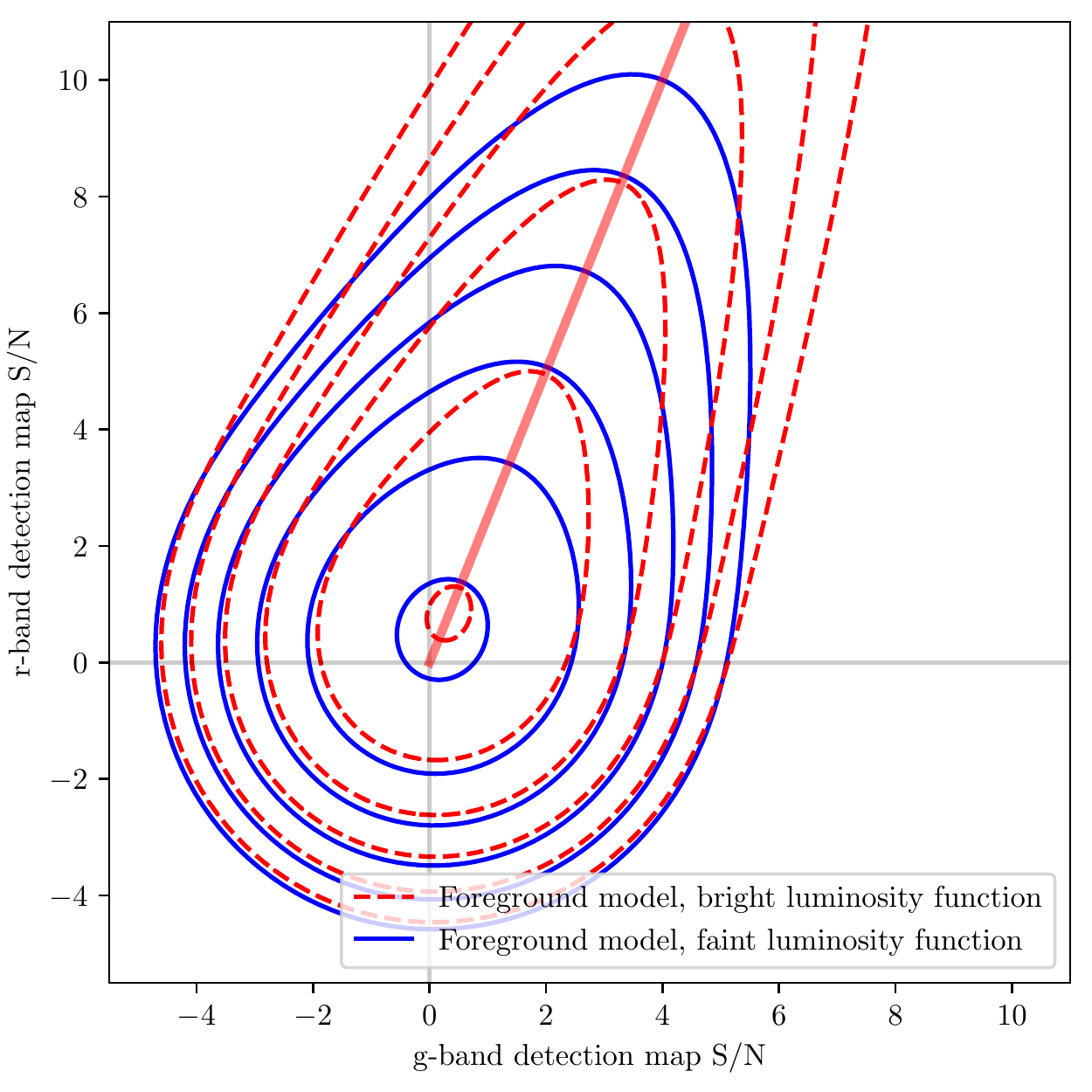}
    \includegraphics[width=0.45\textwidth]{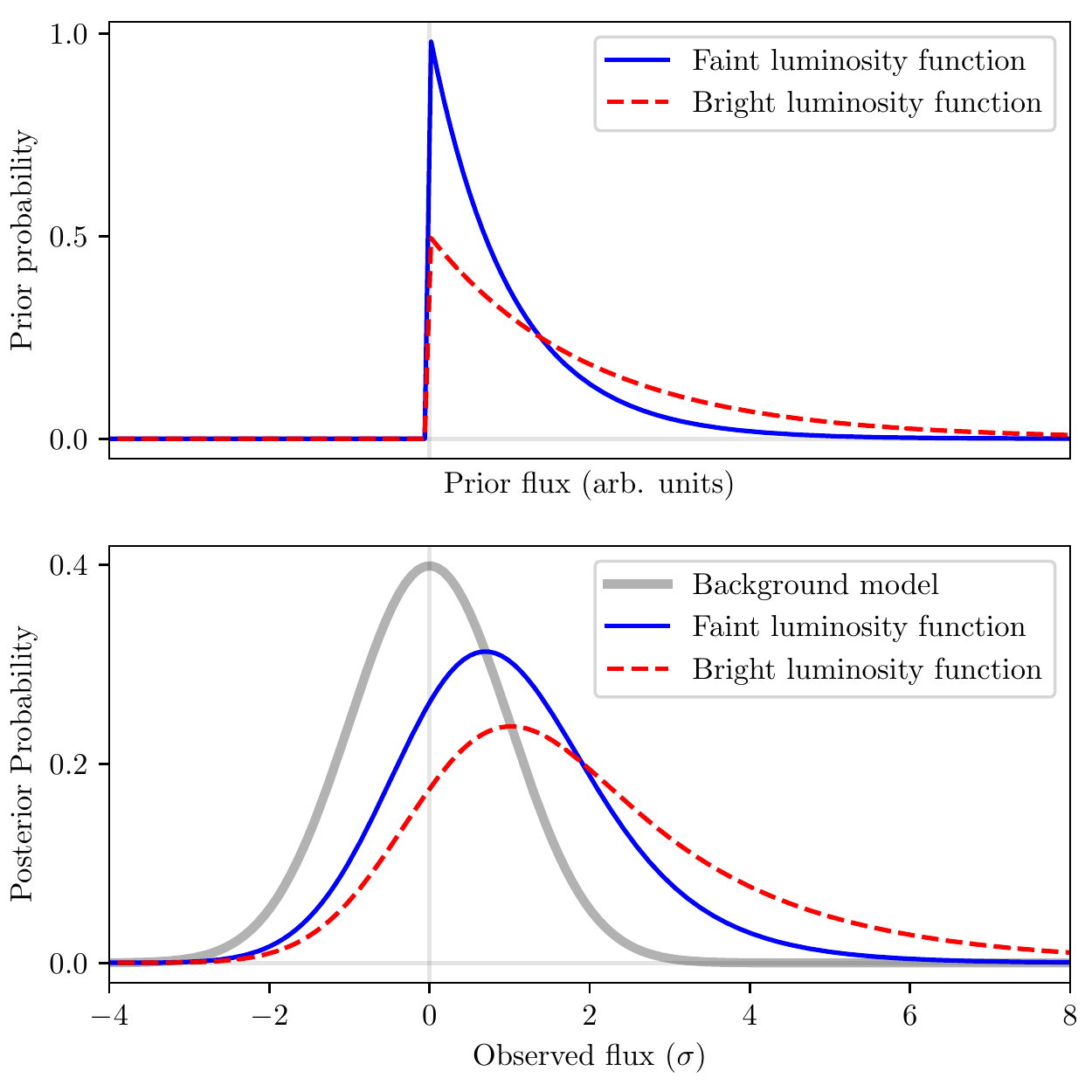}
    \caption{The effect of the flux prior in Bayesian
      SED-matched source detection.  In order to produce closed-form
      integrals, we use an exponential prior on the flux; here
      we show the effect of the exponential scale factor ($\alpha$) on
      the results.  \emph{Left:} Probability contours for the ``Red''
      foreground model, as before, and with a flux prior that
      prefers brighter sources---larger observed fluxes are considered
      more probable.  \emph{Right:} A one-dimensional example showing
      the effect of changing the flux prior by a factor of
      two.  The dashed model places more prior belief in larger
      fluxes, and therefore must produce smaller probabilites at low
      flux levels.  This illustrates that care must be taken with the
      scale factor of the flux prior.
      \label{fig:scaling}
    }
    \end{center}
\end{figure}

As mentioned previously, there is an issue with the flux prior we are
using: it is not scale-free, or rather, it is degenerate with the
choice of $\alpha$.  That is, the posterior probability is sensitive
to the numerical values for the quantity $F$, the ``canonical flux'',
which we have marginalized out of the expression.  We did not put any
restrictions on the SED values $s_{i,j}$; we only stated that the
predicted flux in band $j$ is given by $F \cdot s_{i,j}$.  If we were
to scale up all our $s_{i,j}$ values, this would imply a decrease in
$F$, which would have higher prior probability, leading to larger
posterior probability values.  This problem is illustrated in
\figref{fig:scaling}.  This is obviously undesirable, but thus far we
have not been able to find a scale-free prior that leads to tractable
integrals.

\section{Experiments}

% The experiment here make extensive use of the Legacy Survey code
% infrastructure -- running calibrations of the CP images, and
% producing detection maps.  The processing is described in DECAM.txt,
% and the custom code for this project is in the run-pipe.py file.

We present experiments using data from the Dark Energy Camera
\citep{decam} taken as part of the Supernova program \citep{dessn} of
the Dark Energy Survey \citep{des}.  We selected the deep field ``SN
X3'' near RA,Dec $= (36.45, -4.6)$, which has a large number of
exposures in bands $g$, $r$, $i$ and $z$.  For these experiments, we
use data from bands $g$, $r$ and $i$ only.  We select a set of 25
exposures in each band from the 2016B semester, keeping at most one
exposure per night per band.  The exposure times for each image are
200 seconds in $g$, 400 seconds in $r$, and 360 seconds in $i$.  The
images have a range of seeing and sky transparency values.  The list
of exposures is available in Tables \ref{tab:exposures} and
\ref{tab:exposuresi}.

\begin{figure}
  \begin{center}
    \begin{tabular}{cc}
      Tiny zoom-in & Whole sample \\
      \includegraphics[height=0.4\textwidth]{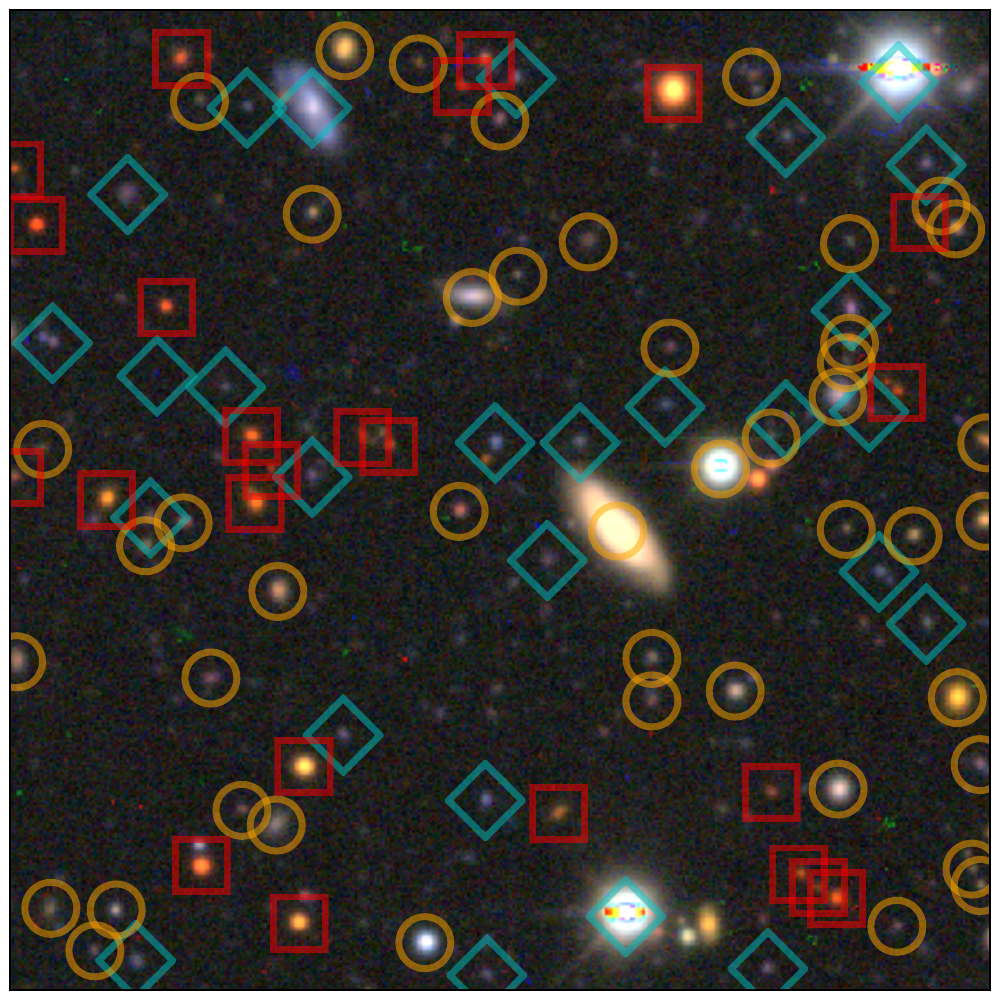}
      &
      \includegraphics[height=0.4\textwidth]{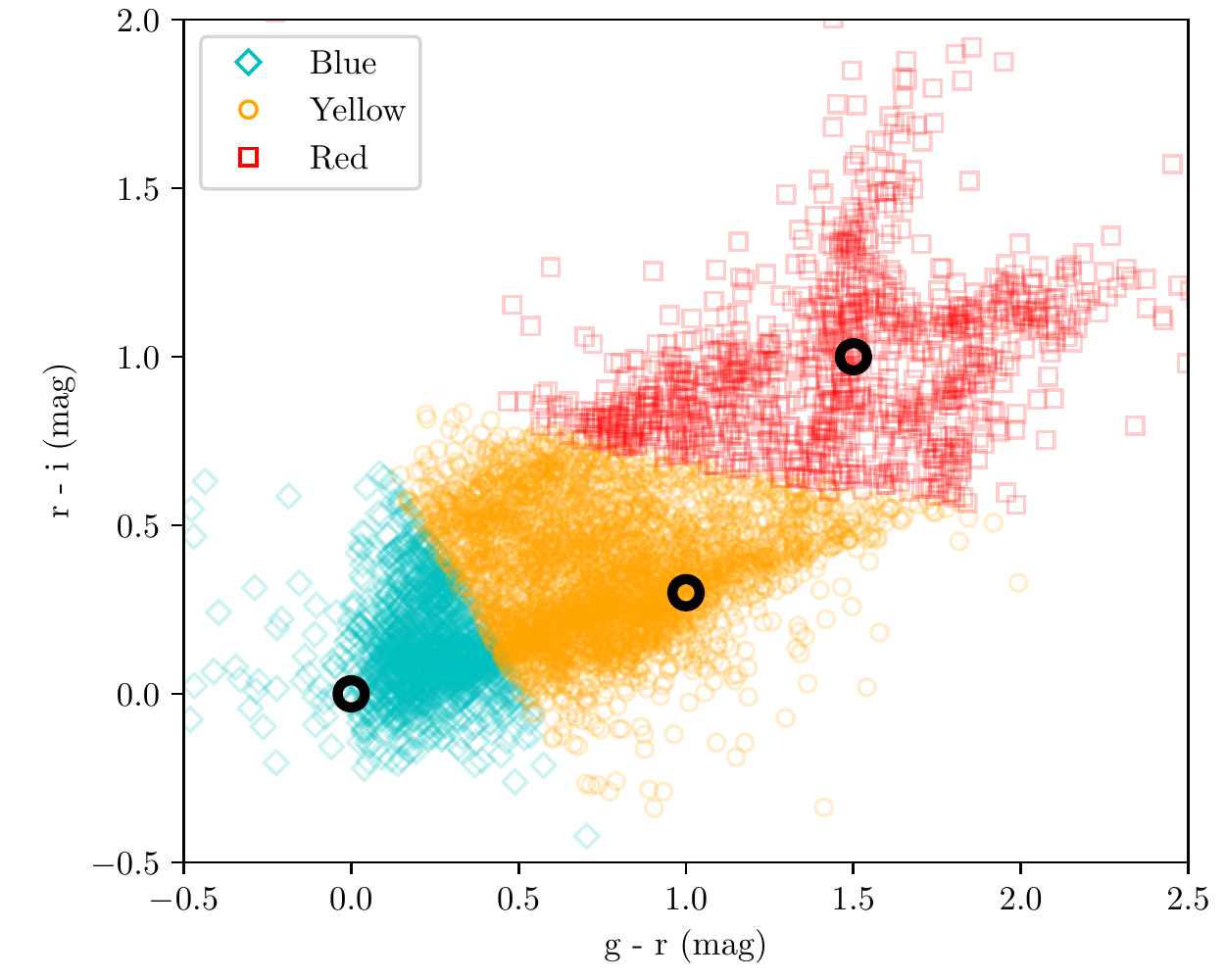}
    \end{tabular}
    \caption{\emph{Left:} A $600 \times 600$-pixel zoom-in of a tiny
      fraction of the DECam data used in this experiment, coadded and
      shown with an RGB color scheme and \emph{arcsinh} stretch.
      Detected sources are marked with symbols indicating which of the
      SED-matched filters yields the strongest detection; for clarity
      here we are using an extremely high detection threshold of $30
      \sigma$.  \emph{Right:}
      The positions in color-space of detected sources, classified by
      the SED-matched filter that yields the strongest detection.  The
      bold circles mark the color to which each filter is tuned.  Each
      SED-matched filter most strongly detects sources nearby in color
      spaces, where the exact dividing line depends on the
      signal-to-noise in the different filters.
      \label{fig:expt}}
  \end{center}
\end{figure}

\begin{figure}
  \begin{center}
    \begin{tabular}{@{}c@{}c@{}c@{}}
      Blue-optimized & Yellow-optimized & Red-optimized \\
      \includegraphics[width=0.33\textwidth]{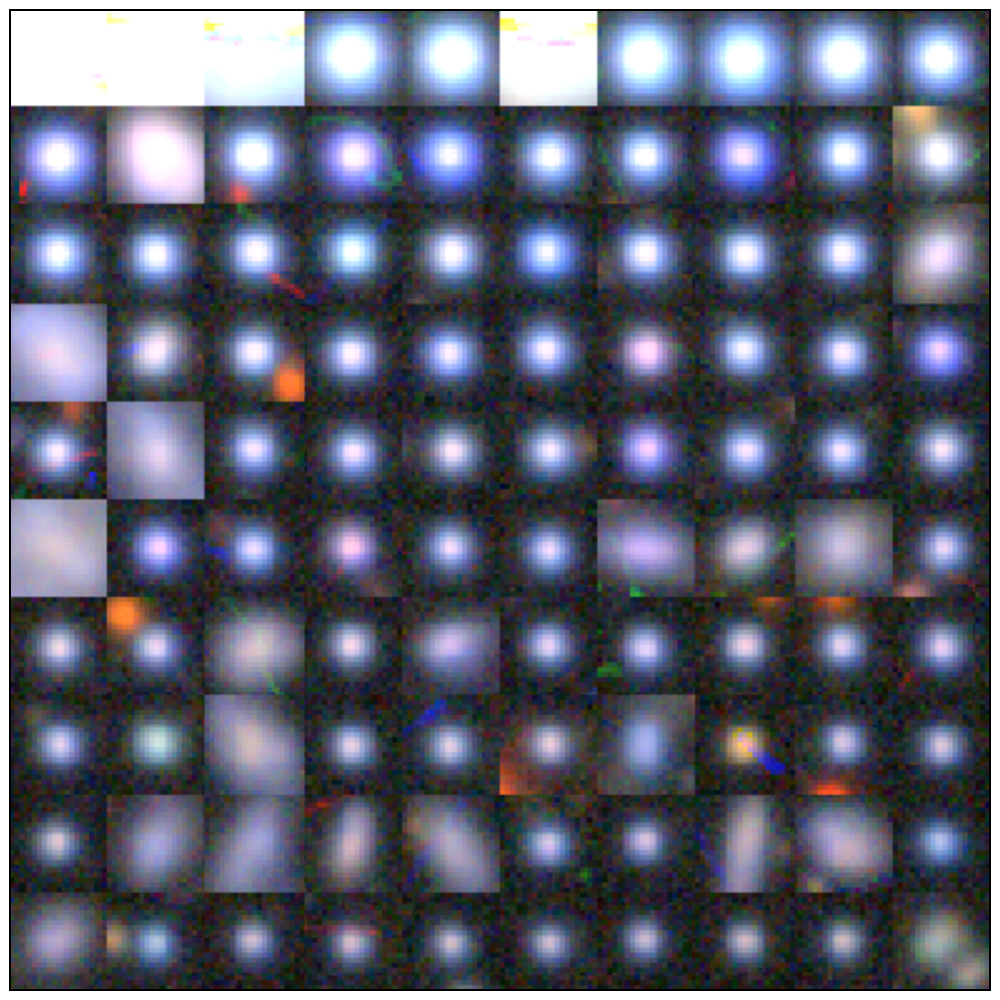} &
      \includegraphics[width=0.33\textwidth]{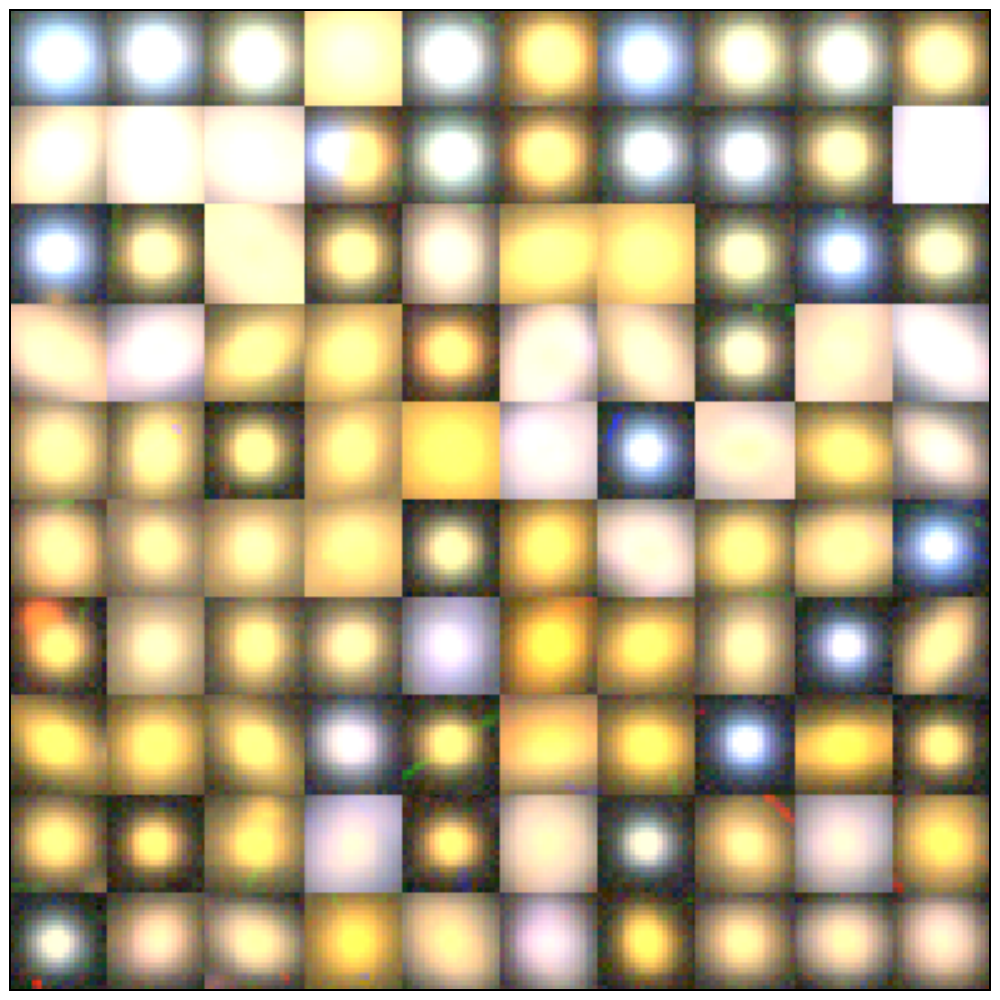} &
      \includegraphics[width=0.33\textwidth]{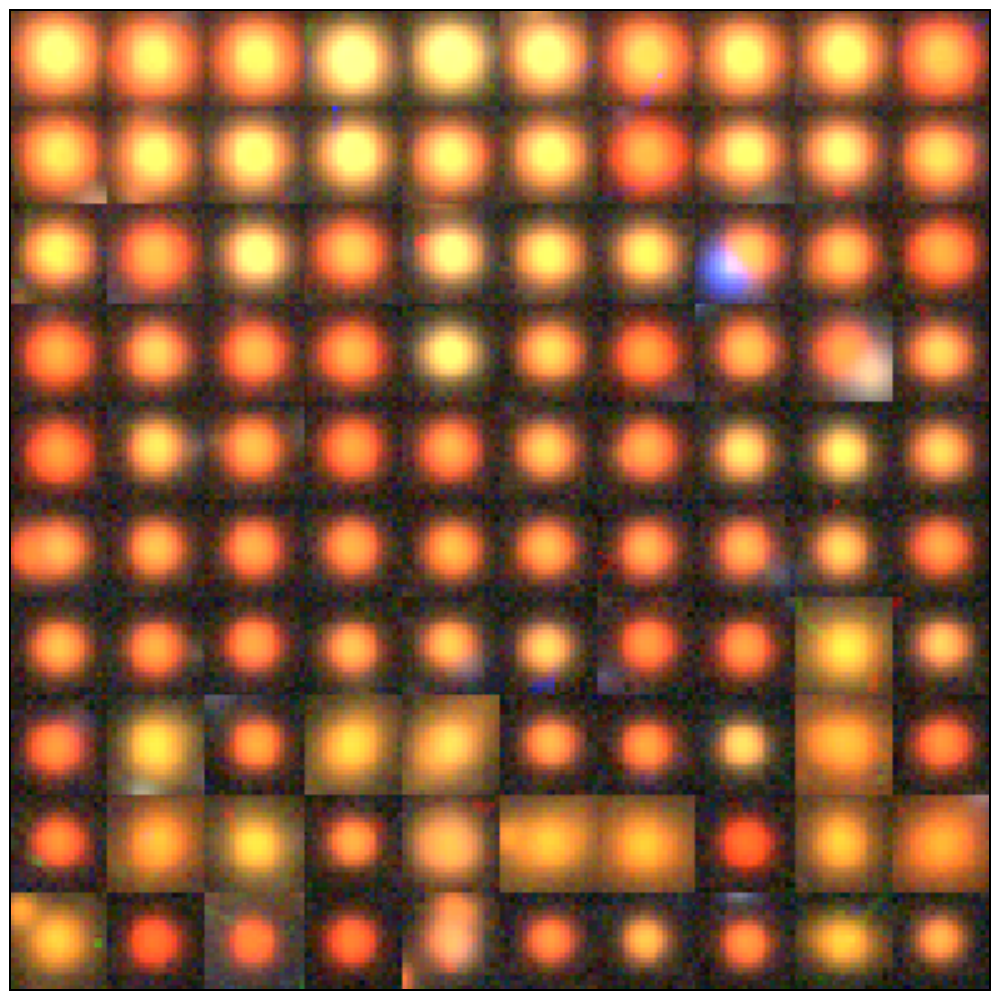}
    \end{tabular}
    \caption{Source that are most strongly detected by each
      SED-matched filter.  \emph{Left:} ``Blue'' SED (matched to color
      $g-r = r-i = 0$).  \emph{Center:} ``Yellow'' SED (matched to
      $g-r = 1$, $r-i = 0.3$).  \emph{Right:} ``Red'' SED (matched to
      $g-r = 1.5$, $r-i = 1$).
    \label{fig:bestsed}}
  \end{center}
\end{figure}

\begin{figure}
  \begin{center}
    \includegraphics[width=0.4\textwidth]{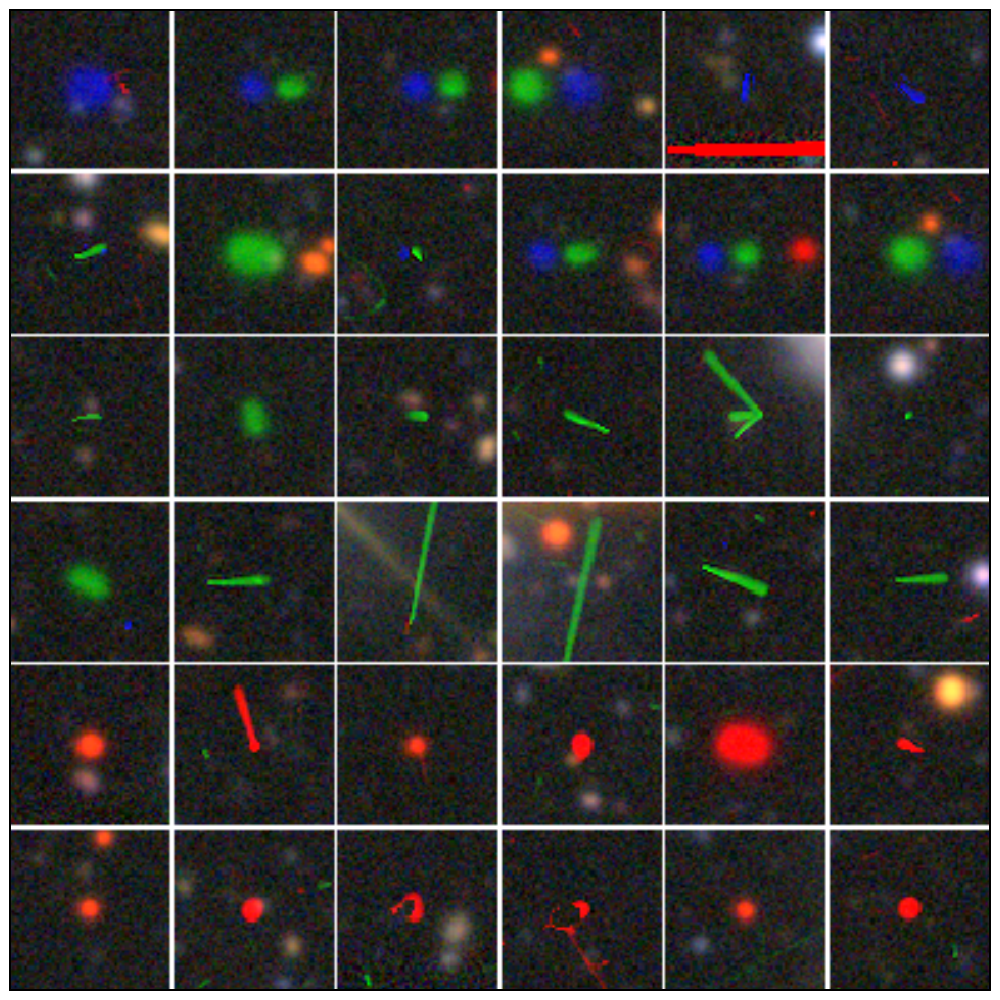}
    \caption{Sources that are detected most strongly in one of the
      single-band SED filters (top row: $g$-band-only; next three
      rows: $r$-band-only; bottom two rows, $i$-band-only).  These
      sources include asteroids (here, 12 observations of 6 known
      asteroids, some visibly elongated during these minutes-long
      exposures), cosmic rays that are not masked by the Community
      Pipeline, and, for $i$-band-only, some extremely red sources.
      \label{fig:oneband}}
  \end{center}
\end{figure}

\begin{figure}
  \begin{center}
    \includegraphics[width=0.9\textwidth]{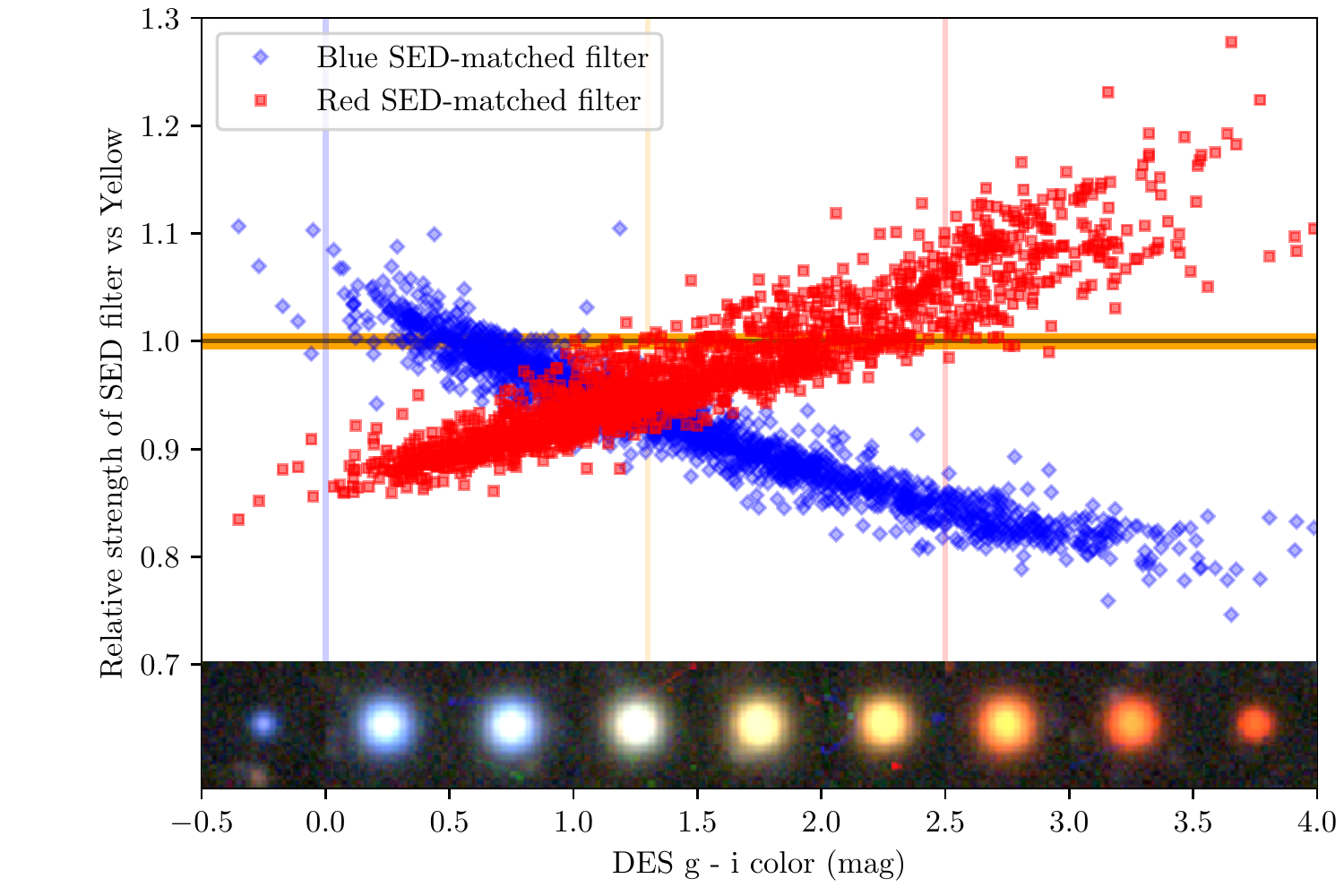}
    \caption{Relative strengths of the different SED-matched detection
      filters.  We match our detected sources with the Dark Energy
      Survey DR1 source database in order to show an accurately
      measured color for each source.  We plot the detection strength
      (sensitivity) of the ``Blue'' and ``Red'' SED-matched filters
      versus the ``Yellow'' filter.  That is, a source with relative
      strength 1.1 in the ``Blue'' filter would be detected at $5.5
      \sigma$ in the ``Blue'' filter if it were detected at $5 \sigma$
      in the ``Yellow'' filter.  As expected, sources with blue $g-i$
      colors are most sensitively detected by the ``Blue'' SED-matched
      filter, sources with intermediate colors are most sensitively
      detected by the ``Yellow'' filter, and red sources are most
      sensitively detected by the ``Red'' SED-matched filter.  The
      $g-i$ colors to which the detections filters are matched are
      marked with vertical lines.  Note that the relative
      sensitivities can be tens of percent, which is very significant
      at low signal-to-noise levels.  The filters are roughly equal at
      the mid-point between the colors for which they are tuned.
      \label{fig:strength}}
  \end{center}
\end{figure}

\begin{figure}
  \begin{center}
    \includegraphics[width=0.32\textwidth]{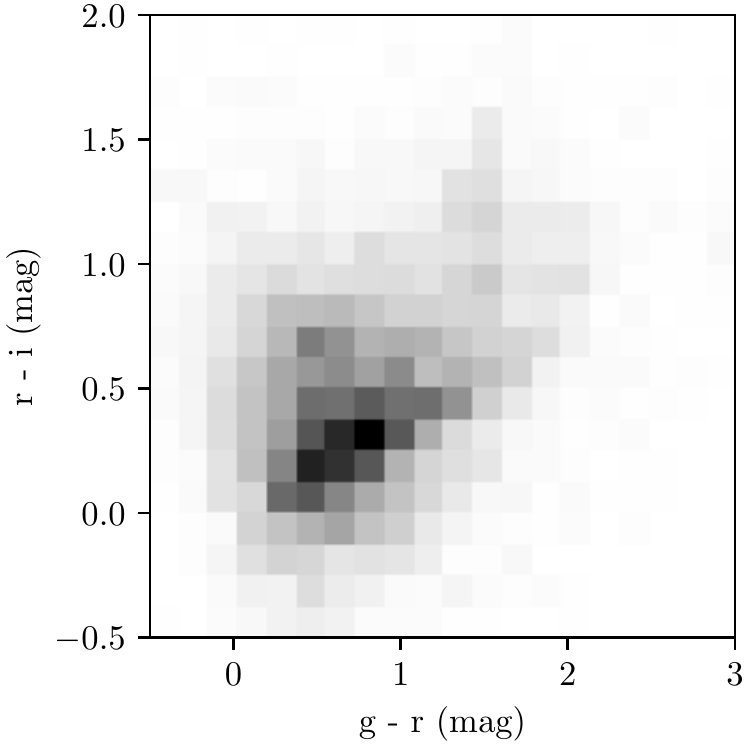}
    \includegraphics[width=0.32\textwidth]{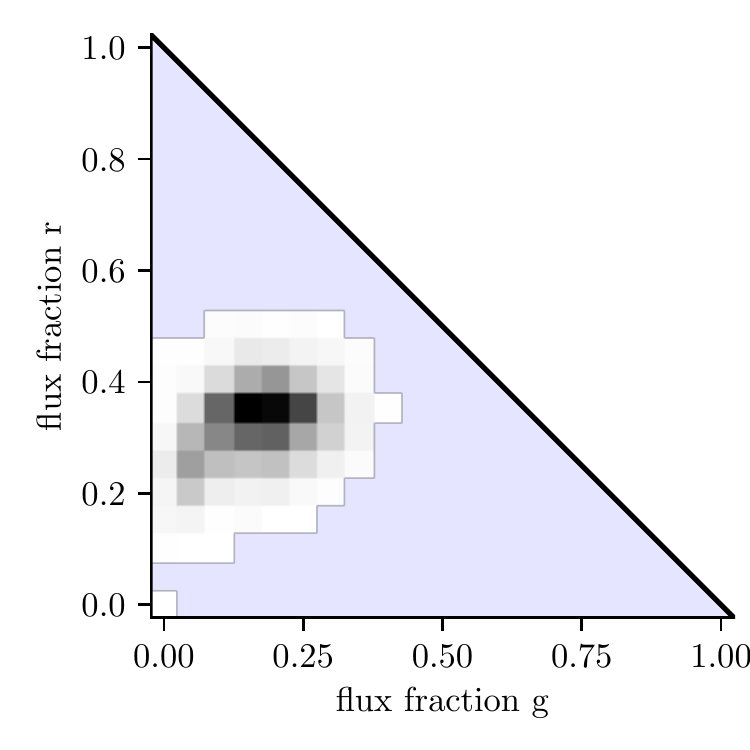}
    \includegraphics[width=0.32\textwidth]{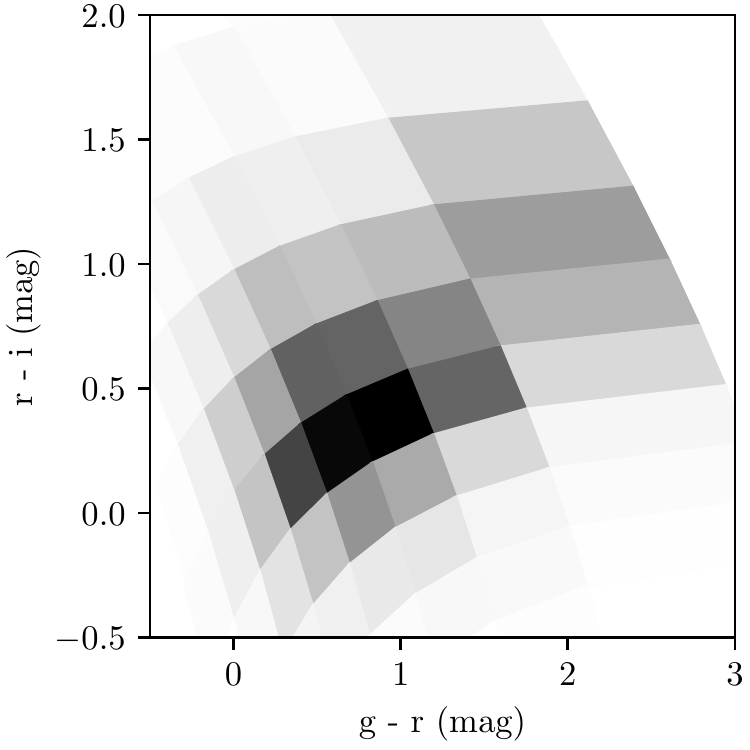}
    \caption{Empirical Bayesian prior for SED-matched detection.
      \emph{Left:} Sources from the Dark Energy Survey catalog, in
      color space.  \emph{Center:} Sources binned in SED space, with
      low-population bins dropped.  The axes are the fraction of flux
      contributed by the $g$ and $r$ bands, respectively.  Implicitly,
      the fraction contributed by the $i$ band is \mbox{1 - ($g$ +
        $r$).}  Empty bins are shown with a light shading.  Notice
      that some bins with $g \sim 0$ are populated; these correspond
      to DES catalog entries with measured $g$ flux consistent with
      zero.  \emph{Right:} Our discrete SED-space binning, projected
      to color space.
      \label{fig:prior}}
  \end{center}
\end{figure}

\begin{figure}
  \begin{center}
    \includegraphics[height=0.4\textwidth]{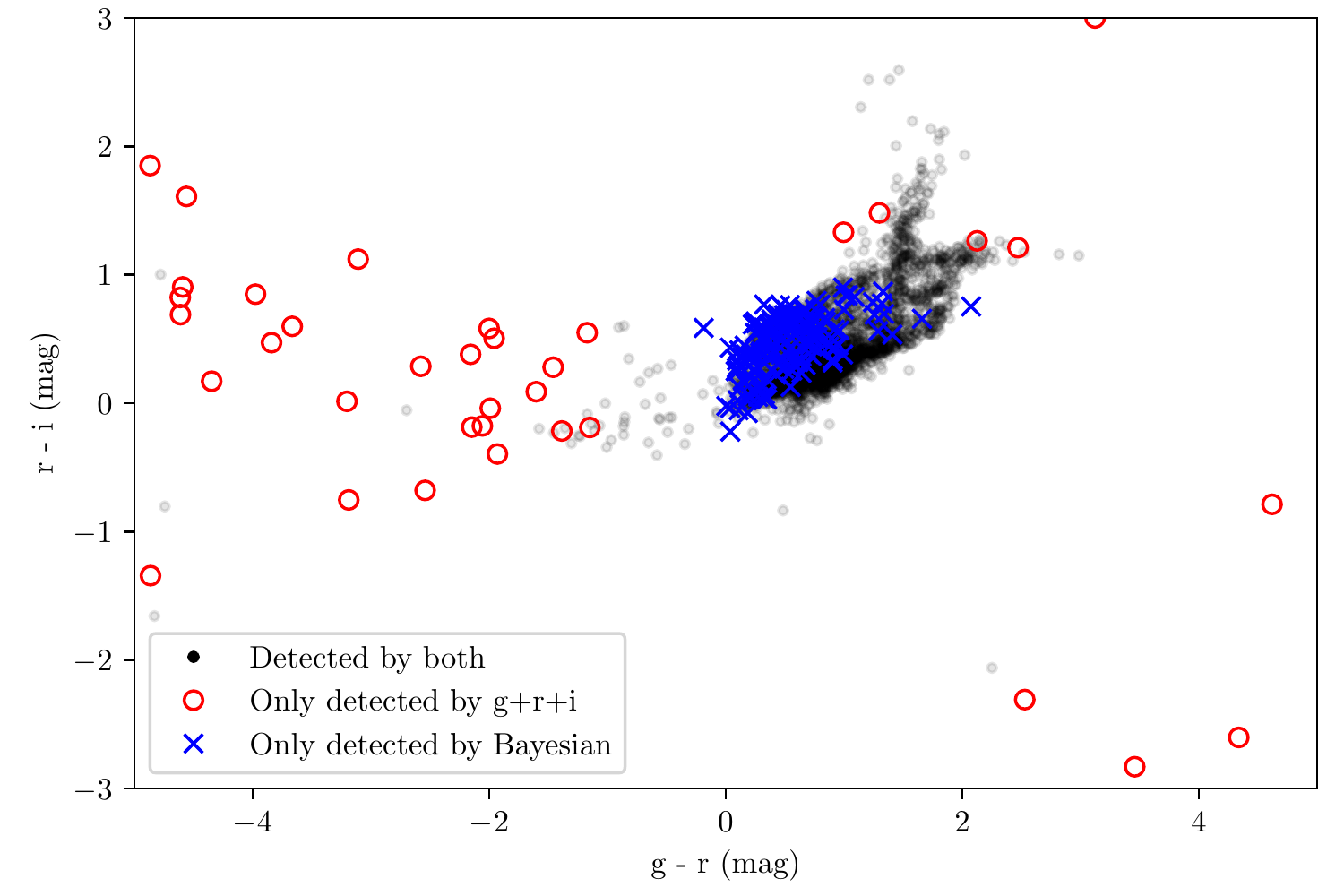}
    \caption{Bayesian SED-matched detection method compared to the
      union of $g$, $r$, and $i$ single-band detection.  The
      color-space locations of sources that are detected by both
      methods are shown as faint points, and sources that are detected
      by only one method or the other are highlighted.  In this
      experiment, the detection thresholds were tuned so that both
      methods produced the same number of detections; this compares
      each method's best detections.  The Bayesian-only detections
      have the colors of real objects, while the $gri$-union
      detections tend to be single-band sources, including cosmic
      rays, asteroids, and very red sources.  Cutouts of these sources
      are shown in \figref{fig:only}.
      \label{fig:bayes-vs-gri}}
  \end{center}
\end{figure}

\begin{figure}[p!]
  \begin{center}
    \begin{tabular}{cc}
      Detected by Bayesian only &
      Detected by $gri$-union only \\
      \includegraphics[width=0.4\textwidth]{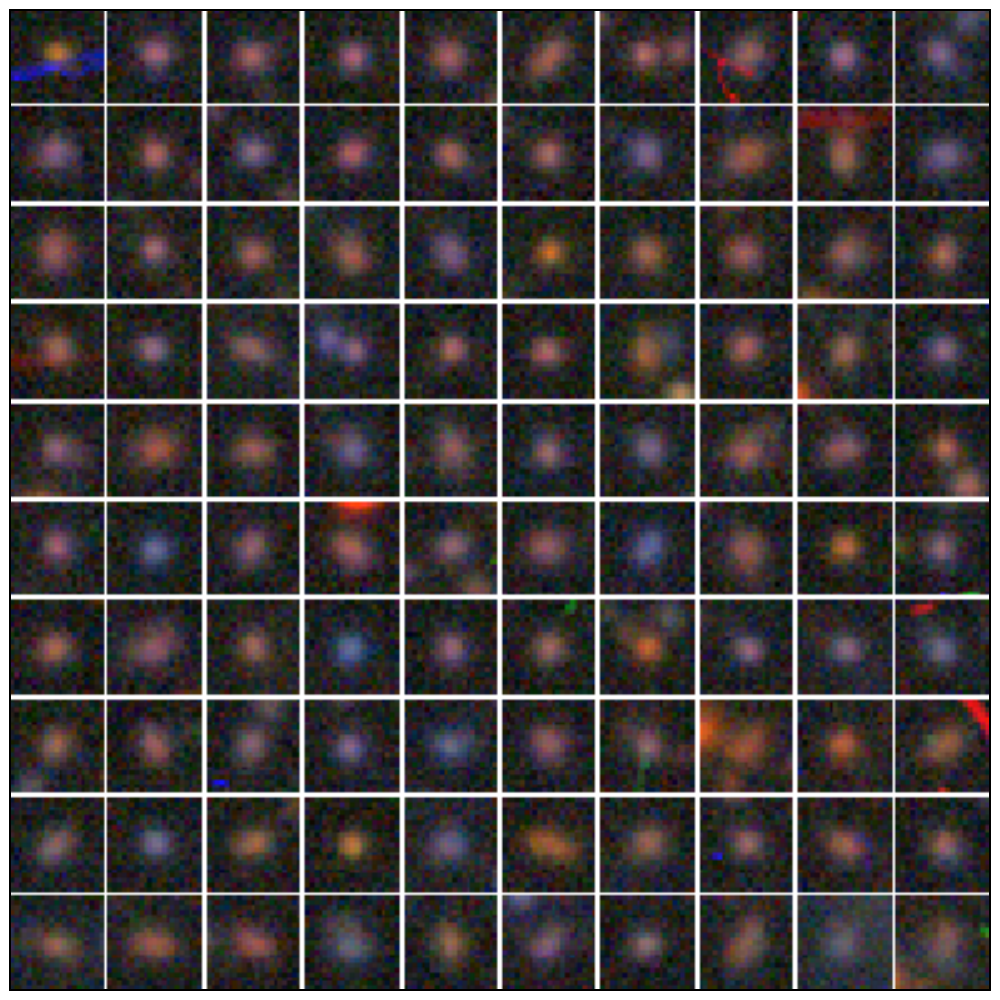}
      %\hspace{0.03\textwidth}
      &
      \includegraphics[width=0.4\textwidth]{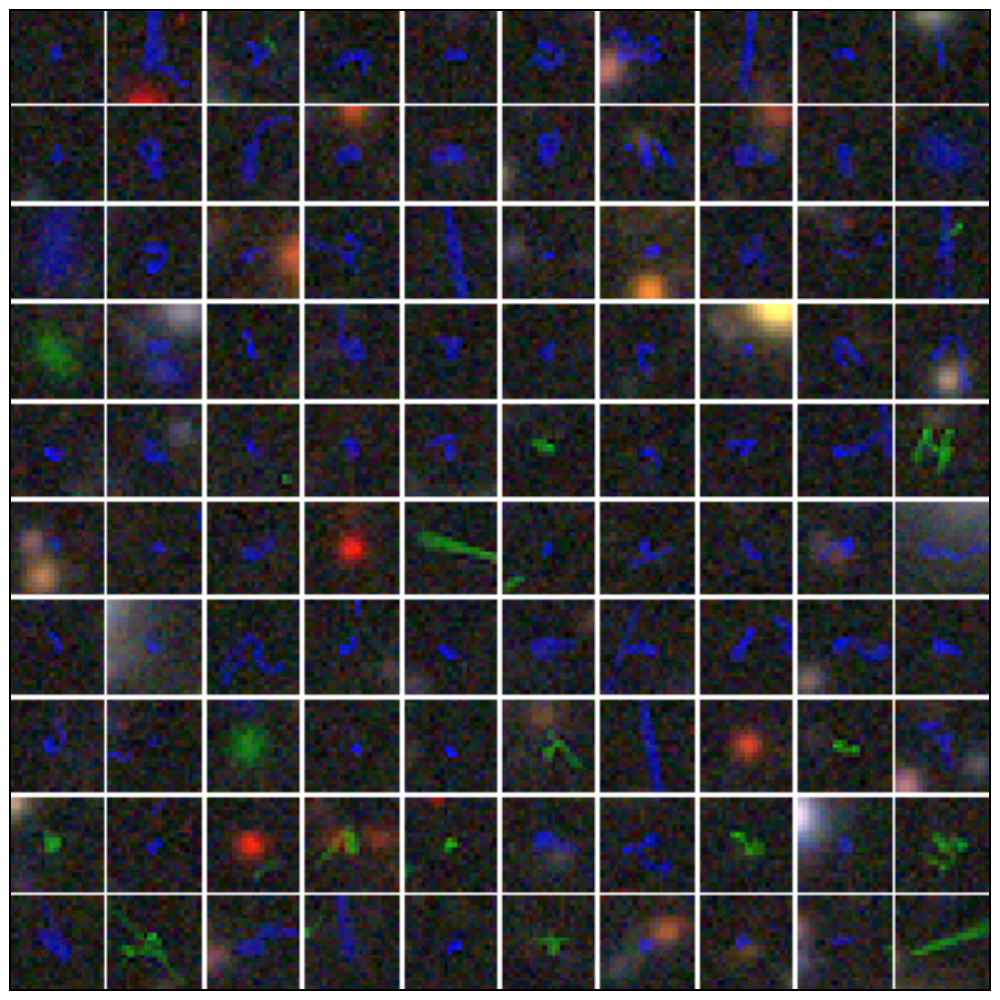}
    \end{tabular}
    \caption{Bayesian SED-matched detection method compared to the
      union of $g$, $r$, and $i$ single-band detections, where each
      method is allowed to detect the same total number of sources.
      \emph{Left:} Examples of sources that are detected by only the
      Bayesian SED-matched detection method.  These have the colors of
      real objects and appear to be all faint galaxies.  \emph{Right:}
      Examples of sources that are detected by only the $gri$-union
      method.  These include cosmic rays,
      asteroids, and very red sources.
      \label{fig:only}}
  \end{center}
\end{figure}

\begin{figure}[p!]
  \begin{center}
    \includegraphics[height=0.4\textwidth]{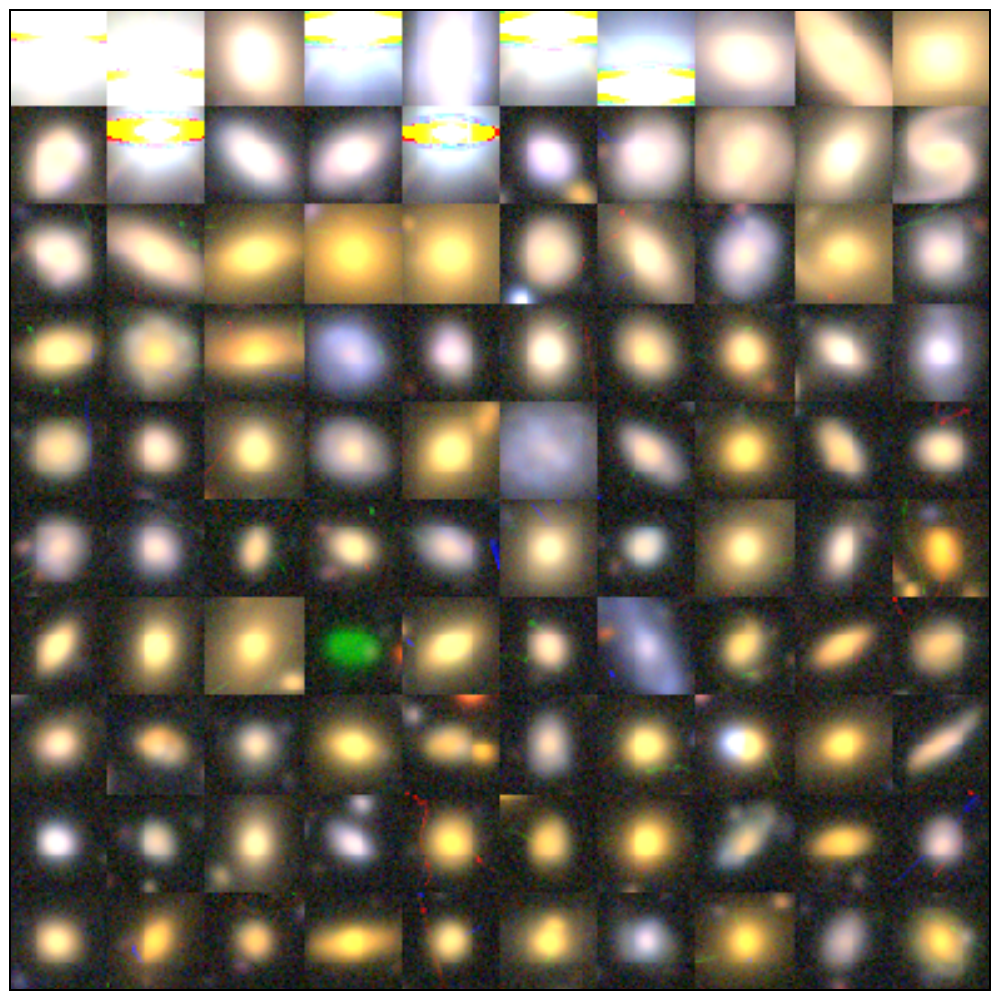}
    \includegraphics[height=0.4\textwidth]{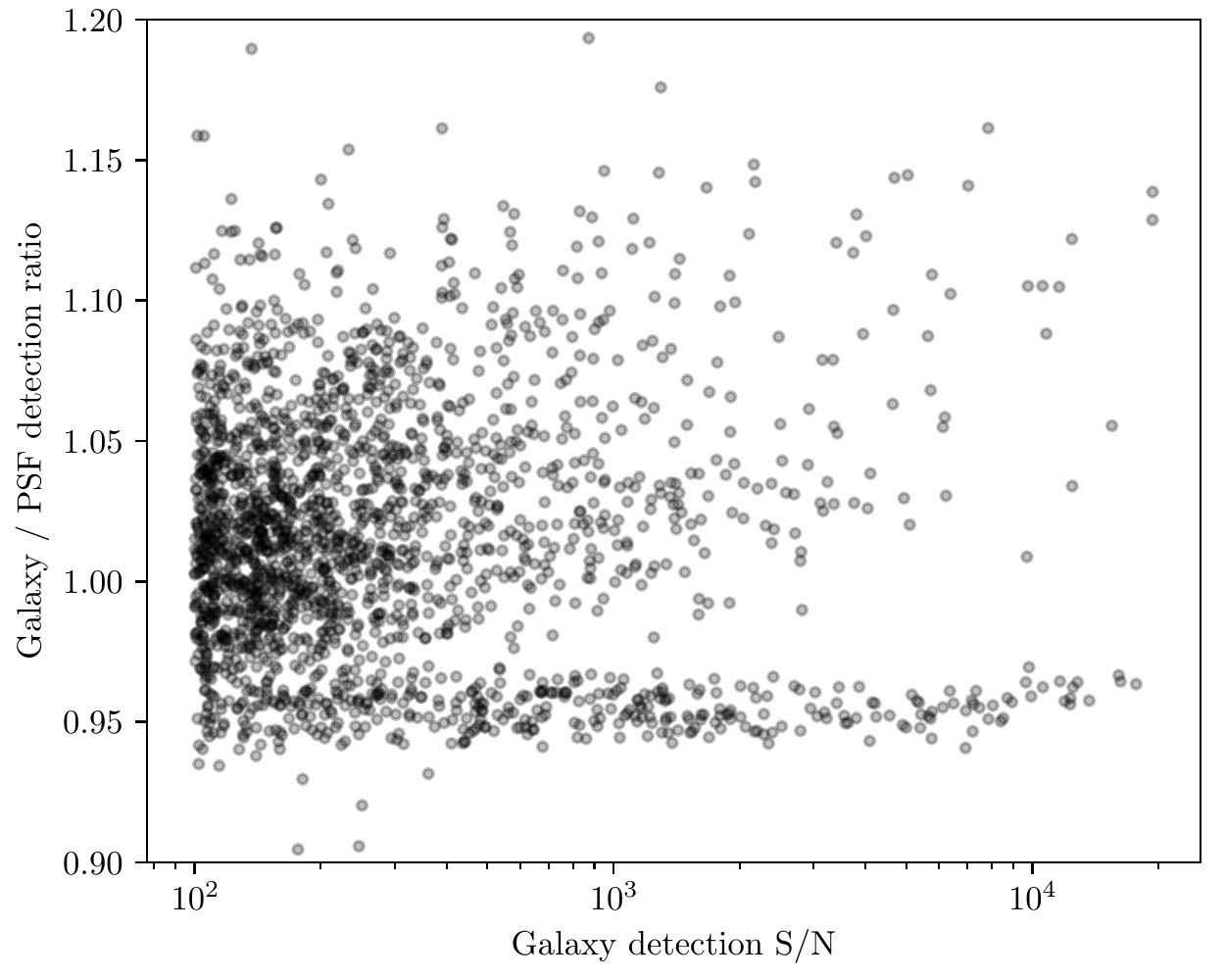}
    \caption{Galaxy detection.  \emph{Left:} Detections from an extended-source detector,
      sorted by the difference in detection strength between the
      ``yellow galaxy'' and ``yellow point source'' detectors.  The
      bright stars are likely included because their saturated cores
      cause difficulty in correctly measuring the signal-to-noise.
      The remaining sources are all galaxies, except for one
      fast-moving asteroid that is spatially extended.  \emph{Right:}
      Relative strength of detection of sources in the extended versus
      PSF detection filters.  The extended-source filter used here is
      a 1-arcsecond FWHM round Gaussian.  There is a clear line of
      sources with a detection ratio around 0.95, corresponding to
      point sources.  Sources with ratios above unity are closer in
      profile to the target profile than to a point source, with the
      largest ratios probably corresponding to sources that are
      approximately equal to the target profile.
      \label{fig:gals}}
  \end{center}
\end{figure}

We use the standard NOAO DECam Community Pipeline \citep{cppipeline}
for calibration of the images, and then compute astrometric and
photometric zeropoints, sky background models and PSF models, using
the DESI Legacy Surveys pipeline \citep{lsoverview},
\textsl{legacypipe}\footnote{Publicly available at
  \niceurl{https://github.com/legacysurvey/legacypipe}.}.
The astrometric
calibration uses Gaia (DR1) as the reference catalog \citep{gaia,
  gaiaDR1}, and the photometric calibration uses the Pan-STARRS DR1
reference catalog \citep{panstarrs}.  The PSF models use PsfEx
\citep{psfex}.

\subsection{SED-matched detection}

We select a 4000-by-4400 pixel region, at the approximate DECam pixel
scale of $0.262$ arcsec/pixel, covering DECam chips N4 and S4 in the
center of the focal plane.  We produce detection maps for each band as
described above, and run several SED-matched filters: a ``Blue''
filter matched to a source with colors $g - r = r - i = 0$; a
``Yellow'' filter matched to $g - r = 1$, $r - i = 0.3$; and a ``Red''
filter matched to $g - r = 1.5$, $r - i = 1$.

For the plots shows here, we select sources in the Yellow filter with
signal-to-noise above $100$, taking as a source the peak pixel within
each connected component of pixels above the threshold; we do not try
to resolve or deblend nearby peaks.  We drop any source near the edge
of the image or where there are fewer than 12 exposures in any of $g$,
$r$, or $i$ bands.  This results in $2092$ detected sources.  For each
source position, we sample the Red and Blue SED-matched maps so that
we can compare the relative sensitivities of the different SED-matched
filters.  We also sample the detection maps in each band at the peak
pixel as an approximate estimate of the flux of the source.  This
assumes all sources are point sources, so underestimates total flux of
galaxies.  An image of the detected sources, and the locations in
color space of sources that are detected by each SED-matched filter
are shown in \figref{fig:expt}.  \figref{fig:bestsed} shows a sample
of sources that are most strongly detected by each of the SED-matched
filters.

We also observe that some sources are most strongly detected in one of
the single-band-only SED-matched filters.  For the $g$-band-only and
$r$-band-only filters, these are almost exclusively image artifacts,
such as cosmic rays that are not masked by the Community Pipeline; and
transient sources, including asteroids.  The $i$-band-only filter can
also detect extremely red sources.  A few examples are shown in
\figref{fig:oneband}.

In \figref{fig:strength} we illustrate how the relative strength of
the different SED-matched filters change with respect to measured
source colors.  We observe differences of tens of percent, which can
lead to a very significant number of additional correctly detected sources at
the lowest signal-to-noise levels.

\subsection{Bayesian SED-matched detection}

We build an empirical ``library'' of SEDs for these bands, which we
use as a prior, by querying the Dark Energy Survey database in our
region of interest.  We convert the \texttt{mag\_auto} measurements
into fluxes, compute the fraction of flux contributed by the $g$, $r$,
and $i$ bands, and histogram these values into a $21 \times 21$ grid.
Keeping bins containing more than $0.1 \%$ of the catalog entries, we
find 63 bins populated, as shown in \figref{fig:prior}.  The prior
includes some SED components that have zero flux in $g$ band, which
correspond to very red sources.

We run the Bayesian detection proceduce described in Equation
\ref{eq:pfg}, computing in log space to avoid numerical overflow.
This takes a few minutes on a single core with our unoptimized
implementation.

In order to show the performance of the Bayesian approach, we will
detect sources at a high threshold and compare against a simple
approach.  For the Bayesian map, we threshold at a log-probability
ratio of $2000$, which yields approximately 3055 sources in good
regions of the image.  For a comparison, we run the same detection
procedure first on our $g$-band map, then on the $r$-band map, and
finally on the $i$-band map, with a detection threshold of $50 \sigma$
for each band, and keeping the union of all detections.  We merge
detected sources if they are within 5 pixels of each other.  This
yields a total of 3201 sources in good regions.  In order to find
sources that are detected by one method and not the other, we first
cut to the brightest 3055 $gri$-union detections (based on maximum
signal-to-noise in the three bands) so that the two lists of sources
are the same length.  We then find sources that are not within an
above-threshold region in the other map.  That is, we ignore Bayesian
detections where the maximum of $g$, $r$, or $i$ signal-to-noise is
above $50$, and we drop $gri$-union detections where the Bayesian
log-probability is above $2000$.  This yields 163 unique Bayesian
detections and 112 unique $gri$-union detections.  As shown in Figures
\ref{fig:bayes-vs-gri} and \ref{fig:only}, the sources detected by
only the Bayesian method have the colors of real sources and appear to
be all real galaxies, while the sources detected by only the $gri$-union
method include a large number of cosmic rays and asteroids, as well as
a number of real very red sources.
If we clip outlier pixels while forming the detection maps, we
can eliminate these cosmic rays before the detection step; the
$gri$-union method then preferentially detects sources with extreme
colors.  Since real images will often contain artifacts
and moving objects, the fact that the $gri$-union method \emph{prefers} these
objects to real sources is problematic; the Bayesian method, on
the other hand, requires sources with such unexpected colors to have
much higher signal than it does for real sources.

%\clearpage
\subsection{Galaxy detection}

While we have focused on the detection of point sources in this paper,
it is a trivial extension to build a detection filter tuned to a
source with a known spatial profile.  We need only correlate the
detection maps for each band by the spatial profile of the source to
be detected, and compute the corresponding change in uncertainty in
the detection maps.  In this experiment, we assume a simple round
Gaussian profile with a FWHM of 1 arcsecond.  After computing
extended-source detection maps for each of the $g$, $r$, and $i$ image
sets, as above, we run the Yellow SED-matched filter and detect
sources as before.  In \figref{fig:gals} we show sources that are more
strongly detected by our extended-source detector than by a
point-source detector.  These include a few bright stars (where the
saturated cores cause errors in estimating the signal-to-noise) but
are largely real galaxies.  Perhaps surprisingly, we find that using a
detection filter tuned to a round Gaussian profile yields only a
few-percent improvement in detection efficiency versus a PSF-tuned
filter, for typical galaxies in DECam images.

\section{Conclusions}

We have reviewed the ubiquitous \emph{matched filter} and shown how it
can be used to detect point sources in collections of astronomical
images.  Our new contribution is to extend the matched filter to SED
space in order to combine images taken through multiple bandpass
filters.  This is a natural extension, but requires specifying the
color of the sources to be detected most sensitively.  A Bayesian
extension of this idea allows us to marginalize over the colors of the
sources to be detected, ideally driven by theoretical or empirical
models of the sources to be detected.  The Bayesian formulation also
requires specifying a prior over the fluxes of sources.  We present an
exponential prior that leads to a closed-form expression for the fully
marginalized Bayesian detection probability; this can be computed
inexpensively.

A pleasing aspect of the matched-filtering approach---including our
extension to SED-matched filtering---is that it can correctly use all
available data.  Images that are known \emph{a priori} to contribute
little information due to their noise levels or because the source is
known to have small flux in that band, are downweighted but not
ignored entirely.  This property will becomes increasingly useful as
multi-wavelength, many-exposure, multi-instrument studies become more
prominent.

\acknowledgments

We thank
Ben Weiner (University of Arizona),
Julianne Dalcanton (University of Washington),
Brad Holden (UCO/Lick Observatories),
Micha Gorelick (Fast Forward Labs),
Robert Lupton, Steve Bickerton, Paul Price, and Craig Loomis (Princeton)
for helpful comments and discussion.

Research at Perimeter Institute is supported in part by the Government
of Canada through the Department of Innovation, Science and Economic
Development Canada and by the Province of Ontario through the Ministry
of Colleges and Universities.

%\bibliographystyle{plain}
%\bibliographystyle{apj}
%\bibliography{detection}

\appendix

\section{Deriving the matched filter}
\label{app:lindet}

Let us assume that there exists a linear filter whose output allows
optimal detection of isolated point sources.  That is, we seek a
(two-dimensional) set of coefficients $a_{\ivec}$ that, when
correlated with the image $I_{\jvec}$, produces a map $M_{\jvec}$
whose peak is the likely location of the point source.

The linear filtering (correlation) operation is
\begin{equation}
M_{\jvec} = \sum_{\iina} a_{\ivec} \, I_{\ivec + \jvec}
\label{eq:detmap1}
\end{equation}
where $\mathcal{A}$ is the support of $\avec$ (integer pixel
positions), and the center of $\avec$ is $\coord{0}{0}$.  We will
demand that the elements of $\avec$ are non-negative and sum to unity.

Inserting \eqnref{eqn:image}, we get
\begin{eqnarray}
M_{\jvec} &\drawnfrom& \sum_{\iina}
  I_k \, a_{\ivec} \, \psfat{\ivec + \jvec - \kvec} + \gaussx{0}{\sigma_1^2}
  \\
&\drawnfrom& \gaussx{ I_k \sum_{\iina} a_{\ivec} \, \psfat{\ivec + \jvec - \kvec}}%
    {\sum_{\iina} a_{\ivec}^2 \, \sigma_1^2}
\end{eqnarray}
and the per-pixel signal-to-noise in the map is
\begin{equation}
%  \signoise_{D_{\jvec}} = \frac{I_k \, \sum_{\iina} a_{\ivec} \, \psfat(\ivec + \jvec - \kvec)}{\sigma_1 \sqrt{\sum_{\iina} a_{\ivec}^2}}
  \snr{M_{\jvec}} = \frac{I_k \, \sum a_{\ivec} \, \psfat{\ivec + \jvec - \kvec}}{\sigma_1 \sqrt{\sum_{\ivec} a_{\ivec}^2}} \quad .
  \label{eq:detmapsn1}
\end{equation}

We want to choose coefficients $a_{\ivec}$ to maximize the 
signal-to-noise at the true pixel position of the source,
$\kvec$.  Rewriting the expression using dot-products and
($\ell_2$) norms, treating the two-dimensional images
$a_{\ivec}$ and $\psfat{\ivec + \jvec - \kvec}$ as vectors indexed by
$\ivec$, we have:
\begin{eqnarray}
  %\signoise_{D_{\jvec}} &=& \frac{I_k \, \avec \cdot \bm{\psfat(\jvec-\kvec)}}{\sigma_1 \sqrt{\avec \cdot \avec}} \\
  \snr{M_{\jvec}} &=& \frac{I_k \, \avec \cdot \bm{\psfat{j-k}}}{\sigma_1 \sqrt{\avec \cdot \avec}} \label{eqn:psfdotprod} \\
 &=& \frac{I_k \, \norm{\avec} \norm{\bm{\psfat{j-k}}} \cos \theta}{\sigma_1 \norm{\avec}} \\
 &=& \frac{I_k \, \norm{\bm{\psfat{j-k}}} \cos \theta}{\sigma_1}
\end{eqnarray}
where $\theta$ is a generalized angle between $\avec$ and $\bm{\psfat{j-k}}$.
At the pixel position of the source, $\kvec$,
\begin{equation}
\snr{M_{\kvec}} = \frac{I_k \, \norm{\bm{\psfat{0}}} \cos \theta}{\sigma_1}
\label{eqn:sndsingle}
\end{equation}
Clearly this is maximized when $\theta = 0$, \ie, when $\avec$ is a
multiple of $\bm{\psfat{0}}$, the PSF evaluated at a grid of integer
pixel positions.  Since we have defined both the PSF and coefficients
$a$ to sum to unity, we find that the optimal linear filter for
detection is given by:
\begin{equation}
%  a_{\ivec} = \psfat{\ivec} \quad ,
\avec = \bm{\psfat{0}} \quad ,
\end{equation}
which means that the operation of \emph{correlating} the image with
its PSF produces a map with optimal signal-to-noise.  Repeating
equation \ref{eq:detmap1}, we have found that the map $M_{\jvec}$ can
be computed by correlating the image with its PSF:
\begin{equation}
M_{\jvec} = \sum_{\iina} \psfat{\ivec} \, I_{\ivec + \jvec} \quad ,
\end{equation}
where, as before, $\mathcal{A}$ is the support of the PSF.
% and
%$\psfat{\ivec} = \psfat{\ivec}$ is an image of the PSF evaluated at
%integer pixel positions $\ivec$.

The signal-to-noise in this map at the true source pixel
position $\kvec$ is
\begin{equation}
\snr{M_{\kvec}} = \frac{I_k \, \norm{\bm{\psfat{}}}}{\sigma_1} \quad .
\end{equation}

\subsection{Optimality}
\label{sec:optsingle}

We compute the variance of the detection map estimator
(\eqnref{eq:detmap}), and show that is equal to the Cram\'er--Rao
bound.  Substituting our image model into the detection map,
\begin{eqnarray}
D_{\jvec} &=& \frac{1}{\psfnorm^2} \sum_{i} \psfat{\ivec} \,
I_{\ivec + \jvec}
\\
& \drawnfrom & \frac{1}{\psfnorm^2} \sum_{i} \psfat{\ivec} \,
\gaussx{F \, \psfat{\ivec + \jvec - \kvec}}{\sigma_1^2}
\end{eqnarray}
and at the true source position, $\jvec = \kvec$;
\begin{eqnarray}
%D_{\jvec} &\drawnfrom& 
%\frac{1}{\psfnorm^2} \sum_{i}
%\gaussx{F \, \psfat{\ivec}^2}{\psfat{\ivec}^2 \sigma_1^2}
%
D_{\kvec} &\drawnfrom& \frac{1}{\psfnorm^2} \gaussx%
{F \, \sum_{i} \psfat{\ivec}^2}%
{\sum_{i} \psfat{\ivec}^2 \sigma_1^2}
\\
D_{\kvec} &\drawnfrom& \gaussx%
{F}{\frac{\sigma_1^2}{\psfnorm^2}}
\end{eqnarray}
so the variance of the estimator is $\var{D} = \frac{\sigma_1^2}{\psfnorm^2}$.

Meanwhile, the Fisher Information for $F$ given pixel values $I_{\jvec}$ is
\begin{eqnarray}
  I(F) &=& -\mathbb{E}_{I_{\jvec}} \left[ \frac{\partial^2 \log P(\{ I_{\jvec} \} | F)}{\partial F^2} \right]
\end{eqnarray}
and with pixel values $I_{\jvec}$ the likelihood is
\begin{eqnarray}
  I_{\jvec} &\drawnfrom& \gaussx{F \psf_{\jvec}}{\sigma_1^2} \\
  P(\{ I_{\jvec} \} | F) &=& \prod_{\jvec} \frac{1}{\sqrt{2 \pi \sigma_1^2}} \exp \left( -\frac{(I_{\jvec} - F \psf_{\jvec})^2}{2 \sigma_1^2} \right) \\
  %\log P(\{ I_{\jvec} \} | F) &=& \sum_{\jvec} \log \frac{1}{\sqrt{2 \pi \sigma_1^2}} + \sum_{\jvec} -\frac{(I_{\jvec} - F \psf_{\jvec})^2}{2 \sigma_1^2} \\
  %\frac{\partial}{\partial F} \log P(\{ I_{\jvec} \} | F) &=& \sum_{\jvec} \frac{(I_{\jvec} - F \psf_{\jvec}) \psf_{\jvec}}{\sigma_1^2} \\
  \frac{\partial^2}{\partial F^2} \log P(\{ I_{\jvec} \} | F) &=& \sum_{\jvec} -\frac{\psf_{\jvec}^2}{\sigma_1^2}
\end{eqnarray}
which is independent of $I_{\jvec}$, so
\begin{eqnarray}
  I(F) &=& \frac{\norm{\psf}^2}{\sigma_1^2}
\end{eqnarray}
from which we see that the estimator $D$ saturates the Cram\'er--Rao bound.

\subsection{Norm of a Gaussian PSF}
\label{app:gaussnorm}
For a Gaussian PSF with standard deviation $\psfw$ pixels,
\begin{eqnarray}\displaystyle
\psf^G(x,y) &=& \frac{1}{2 \pi \psfw^2} \exp{\left(-\frac{x^2}{2 \psfw^2}\right)} \exp{\left(-\frac{y^2}{2 \psfw^2}\right)}
\end{eqnarray}
the norm is% approximately:
\begin{eqnarray}
\norm{\bm{\psf^G}} &=& \sqrt{ \sum_{x} \sum_{y} \left(\frac{1}{2 \pi \psfw^2} \exp{\left(-\frac{x^2}{2 \psfw^2}\right)} \exp{\left(-\frac{y^2}{2 \psfw^2}\right)} \right)^2} \\
%\left(\norm{\psf^G}\right)^2 &\simeq& \iint \frac{1}{4 \pi^2 \psfw^4} \exp{\left(-\frac{x^2}{\psfw^2}\right)} \exp{\left(-\frac{y^2}{\psfw^2}\right)} \mathrm{d}x \, \mathrm{d}y \\
\norm{\bm{\psf^G}} &\simeq& \sqrt{\iint \frac{1}{4 \pi^2 \psfw^4} \exp{\left(-\frac{x^2}{\psfw^2}\right)} \exp{\left(-\frac{y^2}{\psfw^2}\right)} \mathrm{d}x \, \mathrm{d}y} \\
\norm{\bm{\psf^G}} &\simeq& \frac{1}{2 \sqrt{\pi} \psfw}
\end{eqnarray}
so the \detmap\ has signal-to-noise at the true source position $\kvec$,
\begin{equation}
\snr{D_{\kvec}^G} = \frac{I_k}{2 \sqrt{\pi} \psfw \sigma_1 } \quad .
\label{eqn:sndsinglegauss}
\end{equation}

Note, however, that we have defined the point-spread function
$\psf(\cdot)$ to be the \emph{pixel-convolved} response, so it cannot
be exactly Gaussian if the pixel response is assumed to be a boxcar
function.  In practice, however, a two-dimensional Gaussian with
variance $v^2$ correlated with a two-dimensional boxcar function is
well approximated by a Gaussian with variance $v^2 + \frac{1}{12}$, as
long as $v \gtrsim \frac{1}{2}$.

% figure showing this?

\subsection{Why not signal-to-noise-squared?}
In correlating the image with the PSF, it looks like the
\detmap\ weights pixels by their signal-to-noise, rather than
signal-to-noise \emph{squared}.  This apparent conflict can be
resolved by scaling the pixel values so that each pixel is an estimate
of the same quantity.  That is, we want to estimate the total source
counts $F$, but the pixels contain estimates of the source counts
scaled by the PSF, $F \psf$; we must undo this scaling by multiplying
the pixels by $1/\psf$.

Given a source at position $\kvec$, we define the image $K$ whose
pixels each contain an estimate of the total source counts:
\begin{eqnarray}
  K_{\jvec} &=& \frac{1}{\psfat{\jvec-\kvec}} S_{\jvec}  \\
  K_{\jvec} &\drawnfrom& \frac{1}{\psfat{\jvec-\kvec}} \, \gaussian{I_k \, \psfat{\jvec-\kvec}\, , \, \sigma_1^2} \\
  K_{\jvec} &\drawnfrom& \gaussian{I_k \, , \, \frac{\sigma_1^2}{\psfat{\jvec-\kvec}^2}} \quad .
\end{eqnarray}

The signal-to-noise remains the same, since we have just scaled the
values:
\begin{eqnarray}
\snr{K_{\jvec}} &=& \frac{I_k \, \psfat{\jvec-\kvec}}{\sigma_1} \\
\snr{K_{\jvec}} &=& \snr{S_{\jvec}} \quad .
\end{eqnarray}

As before, the \detmap\ pixels are a linear combination of the
(shifted) pixels of the $K$ image with weights $b_{\ivec}$:
\begin{eqnarray}
D_{\jvec}^\star &=& \sum_{\ivec} b_{\ivec} \, K_{\ivec+\jvec} \\
&\drawnfrom& \sum_{\ivec} b_{\ivec} \, \gaussian{I_k \,,\, \frac{\sigma_1^2}{\psfat{\ivec+\jvec-\kvec}^2}} \\
&\drawnfrom& \gaussian{I_k \sum_{\ivec} b_{\ivec} \,,\, \sum_{\ivec} \frac{b_{\ivec}^2 \sigma_1^2}{\psfat{\ivec+\jvec-\kvec}^2}}
\end{eqnarray}
and the signal-to-noise in that \detmap\ at pixel $\kvec$ is
\begin{eqnarray}
\snr{D_{\kvec}^\star} &=& \frac{I_k \sum_{\ivec} b_{\ivec}}{\sigma_1 \sqrt{\sum_{\ivec} \frac{b_{\ivec}^2}{\psfat{\ivec}^2}}}
\end{eqnarray}
which is maximized by setting the $b_{\ivec}$
\begin{equation}
b_{\ivec} \propto \psfat{\ivec}^2 \quad :
\end{equation}
proportional to the signal-to-noise \emph{squared}, as expected.

% Derivative is
% d\snr{D_k^\star}/db_k = 
%  \[ \frac{f}{\sigma} [ \frac{1}{\sqrt{\sum_j \frac{b_j^2}{\psfat{j}^2}}}
%                       - \frac{(\sum_j b_j) b_k}{\psfat{j}^2 (\sum_j \frac{b_j^2}{\psfat{j}^2})^{3/2}} ]
% \]
%
% And through some seemingly circular math you get to:
% b_k = \psfat{k}^2 * ( \sum_j (b_j^2 / \psfat(j)^2) ) / (sum_j (b_j))
%
% (you can remove the b_k term from the sums over j if you want, with no effect.)
%
% So b_k = \alpha \psfat{k}^2; substituting that you get:
%
% b_k = \psfat{k}^2 * ( \sum_j ( (\psfat{j}^2 \alpha)^2 / \psfat{j}^2 ) / (sum_j (\psfat{j}^2 \alpha))
%     = \psfat{k}^2 * ( \alpha^2 \sum_j \psfat{j}^2 ) / (\alpha \sum_j(\psfat{j}^2))
%     = \psfat{k}^2 * \alpha

\section{Multi-image detection}
\label{app:multidet}

\subsection{Optimality}
\label{app:multiopt}
As before, we will show that the estimator $F^{\star}$ in
Equation \ref{eq:onebandmap} saturates the Cram\'er--Rao bound for $F$.

We will consider two images, $A$ and $B$, with PSFs $\psi$ and $\phi$,
respectively, and calibration factors $\kappa_A$ and $\kappa_B$ that
scale image units to flux units.  Per-pixel noise in the two images
will be $\sigma_A$ and $\sigma_B$.  We will assume that the pixel
grids are aligned so that no resampling is necessary.

Given all this, the pixel value for images $A$ and $B$ are drawn from
the distributions
\begin{eqnarray}
  A & \drawnfrom & \gaussx{\frac{F}{\kappa_A} \psi_k}{\sigma_A^2} \\
  B & \drawnfrom & \gaussx{\frac{F}{\kappa_B} \phi_k}{\sigma_B^2}
  \quad .
\end{eqnarray}
The Fisher Information is
\begin{eqnarray}
  I(F) &=& -\mathbb{E}_{A,B} \left[ \frac{\partial^2 \log P(\{ A,B \} | F)}{\partial F^2} \right]
\end{eqnarray}
and assuming that images $A$ and $B$ are statistically independent,
$P(A,B | F) = P(A|F) P(B|F)$.  Following the analysis in
\ref{sec:optsingle}, we find that
\begin{eqnarray}
  I(F) &=& \frac{\norm{\psi}^2}{\kappa_A^2 \sigma_A^2} +
  \frac{\norm{\phi}^2}{\kappa_B^2 \sigma_B^2}
\end{eqnarray}
which equals the variance of the $F^{\star}$ estimator,
$\sigma_{F^{\star}}^2$ as given in Equation \ref{eq:onebandstd}.
Therefore, the estimator saturates the Cram\'er--Rao bound.

\subsection{Experiments}

\begin{table}
  \begin{center}
    \begin{scriptsize}
    \begin{tabular}{cccccc}
      \hline
      Filter & Exposure number & Date & Exposure time (s) &
      Seeing (arcsec) & Depth ($5 \sigma$ point source) \\
      \hline
      g & 563982 & 2016-08-14 & 200 & 1.28 & 24.65 \\
      g & 566968 & 2016-08-24 & 200 & 1.79 & 23.02 \\
      g & 567422 & 2016-08-25 & 200 & 1.72 & 23.55 \\
      g & 569591 & 2016-08-31 & 200 & 1.55 & 24.39 \\
      g & 571049 & 2016-09-05 & 200 & 1.67 & 24.29 \\
      g & 573546 & 2016-09-11 & 200 & 1.42 & 24.49 \\
      g & 574702 & 2016-09-14 & 200 & 2.17 & 22.75 \\
      g & 575794 & 2016-09-22 & 200 & 1.20 & 24.17 \\
      g & 577432 & 2016-09-26 & 200 & 1.55 & 24.47 \\
      g & 579874 & 2016-10-02 & 200 & 1.37 & 24.51 \\
      g & 582140 & 2016-10-09 & 200 & 1.61 & 23.56 \\
      g & 584106 & 2016-10-20 & 200 & 1.64 & 24.05 \\
      g & 585888 & 2016-10-25 & 200 & 2.00 & 24.11 \\
      g & 588620 & 2016-11-02 & 200 & 1.39 & 24.43 \\
      g & 591449 & 2016-11-09 & 200 & 1.47 & 23.53 \\
      g & 593383 & 2016-11-17 & 200 & 1.53 & 24.39 \\
      g & 595093 & 2016-11-22 & 200 & 1.45 & 24.48 \\
      g & 596474 & 2016-11-26 & 200 & 1.07 & 24.80 \\
      g & 598232 & 2016-12-01 & 200 & 1.96 & 23.91 \\
      g & 600846 & 2016-12-08 & 200 & 1.20 & 23.69 \\
      g & 601468 & 2016-12-17 & 200 & 1.32 & 23.41 \\
      g & 603288 & 2016-12-22 & 200 & 1.55 & 24.35 \\
      g & 604684 & 2016-12-28 & 200 & 1.82 & 24.26 \\
      g & 605946 & 2017-01-03 & 200 & 1.75 & 24.11 \\
      g & 609567 & 2017-01-17 & 200 & 1.17 & 24.14 \\
      \hline
      r & 563978 & 2016-08-14 & 400 & 1.37 & 24.51 \\
      r & 566976 & 2016-08-24 & 400 & 1.58 & 23.57 \\
      r & 567426 & 2016-08-25 & 400 & 1.53 & 23.99 \\
      r & 569613 & 2016-08-31 & 400 & 1.18 & 24.82 \\
      r & 571060 & 2016-09-05 & 400 & 1.64 & 24.51 \\
      r & 573562 & 2016-09-11 & 400 & 1.34 & 24.19 \\
      r & 574711 & 2016-09-14 & 400 & 1.47 & 23.82 \\
      r & 575798 & 2016-09-22 & 400 & 1.07 & 24.33 \\
      r & 576542 & 2016-09-24 & 400 & 1.67 & 24.44 \\
      r & 578740 & 2016-09-29 & 400 & 1.54 & 24.46 \\
      r & 580295 & 2016-10-03 & 400 & 1.62 & 24.34 \\
      r & 582423 & 2016-10-10 & 400 & 1.07 & 24.12 \\
      r & 584144 & 2016-10-20 & 400 & 1.55 & 23.91 \\
      r & 585892 & 2016-10-25 & 400 & 1.90 & 24.15 \\
      r & 588624 & 2016-11-02 & 400 & 1.25 & 24.61 \\
      r & 591453 & 2016-11-09 & 400 & 1.16 & 24.30 \\
      r & 593076 & 2016-11-16 & 400 & 1.15 & 23.52 \\
      r & 593387 & 2016-11-17 & 400 & 1.35 & 24.64 \\
      r & 595359 & 2016-11-23 & 400 & 1.85 & 24.20 \\
      r & 597239 & 2016-11-28 & 400 & 1.84 & 24.33 \\
      r & 598940 & 2016-12-03 & 400 & 1.43 & 24.46 \\
      r & 600880 & 2016-12-08 & 400 & 2.67 & 23.26 \\
      r & 601776 & 2016-12-18 & 400 & 1.20 & 24.98 \\
      r & 604334 & 2016-12-25 & 400 & 1.01 & 25.05 \\
      r & 605255 & 2016-12-30 & 400 & 1.50 & 24.64 \\
      \hline
    \end{tabular}
    \caption{Exposures in $g$ and $r$ bands from the Dark Energy
      Camera used in the experiments.\label{tab:exposures}}
    \end{scriptsize}
  \end{center}
\end{table}

\begin{table}
  \begin{center}
    \begin{scriptsize}
    \begin{tabular}{cccccc}
      \hline
      Filter & Exposure number & Date & Exposure time (s) &
      Seeing (arcsec) & Depth ($5 \sigma$ point source) \\
      \hline
      i & 563972 & 2016-08-14 & 360 & 1.77 & 23.49 \\
      i & 566980 & 2016-08-24 & 360 & 1.35 & 23.61 \\
      i & 567442 & 2016-08-25 & 360 & 1.05 & 24.20 \\
      i & 567867 & 2016-08-26 & 360 & 1.04 & 24.24 \\
      i & 570175 & 2016-09-02 & 360 & 1.81 & 23.10 \\
      i & 571447 & 2016-09-06 & 360 & 1.54 & 23.23 \\
      i & 573865 & 2016-09-12 & 360 & 1.66 & 23.01 \\
      i & 574727 & 2016-09-14 & 360 & 1.80 & 22.73 \\
      i & 575802 & 2016-09-22 & 360 & 1.09 & 23.13 \\
      i & 576546 & 2016-09-24 & 360 & 1.41 & 23.33 \\
      i & 579449 & 2016-10-01 & 360 & 1.15 & 23.41 \\
      i & 581861 & 2016-10-08 & 360 & 1.47 & 23.03 \\
      i & 584166 & 2016-10-20 & 360 & 1.27 & 23.01 \\
      i & 585960 & 2016-10-25 & 360 & 1.73 & 22.89 \\
      i & 588628 & 2016-11-02 & 360 & 1.20 & 23.93 \\
      i & 591457 & 2016-11-09 & 360 & 1.06 & 24.00 \\
      i & 593080 & 2016-11-16 & 360 & 1.03 & 23.51 \\
      i & 595056 & 2016-11-22 & 360 & 1.81 & 23.75 \\
      i & 596517 & 2016-11-26 & 360 & 1.61 & 24.00 \\
      i & 598236 & 2016-12-01 & 360 & 1.38 & 23.88 \\
      i & 600850 & 2016-12-08 & 360 & 1.04 & 24.12 \\
      i & 601780 & 2016-12-18 & 360 & 1.12 & 24.44 \\
      i & 604338 & 2016-12-25 & 360 & 1.14 & 24.23 \\
      i & 605266 & 2016-12-30 & 360 & 1.67 & 23.90 \\
      i & 607844 & 2017-01-09 & 360 & 1.34 & 22.65 \\
      \hline
    \end{tabular}
    \caption{Exposures in $i$ band from the Dark Energy Camera used in
      the experiments.\label{tab:exposuresi}}
    \end{scriptsize}
  \end{center}
\end{table}

\end{document}